\documentclass[aps,prd,twocolumn,tightenlines,superscriptaddress,nofootinbib,preprintnumbers]{revtex4-2}

\usepackage{amsmath}
\usepackage{amssymb}
\usepackage{latexsym}
\usepackage{graphicx}
\usepackage{slashed}
\usepackage{pifont}
\usepackage{color}
\usepackage{multirow}
\usepackage[usenames,dvipsnames]{xcolor}
\usepackage[caption=false]{subfig}
\usepackage[colorlinks=True, citecolor=blue, urlcolor=blue, linkcolor=blue]{hyperref}
\usepackage{cleveref}
\usepackage{orcidlink}
\usepackage{float}
\usepackage{booktabs}
\usepackage[normalem]{ulem}

\hypersetup{colorlinks=true,citecolor=[rgb]{0,0,0.8},
linkcolor=[rgb]{0,0,0.8},urlcolor=[rgb]{0,0,0.8}}

\newcommand{\JWX}[1]{\textcolor{black}{#1}}
\newcommand{\JW}[1]{\textcolor{black}{#1}}

\newcommand{\XY}[1]{\textcolor{black}{#1}}

\newcommand{\nn}{\nonumber}

\newcommand{\CNOT}{\textrm{CNOT}}

\begin{document}

\title{Local Thermalization of SU(2) Lattice Gauge Fields on Quantum Computers}

\author{Jiunn-Wei Chen}
\email{jwc@phys.ntu.edu.tw}
\affiliation{Department of Physics and Center for Theoretical Physics,
National Taiwan University, Taipei, Taiwan 106}
\affiliation{Physics Division, National Center for Theoretical Sciences, Taipei 10617, Taiwan}
\affiliation{InQubator for Quantum Simulation (IQuS), Department of Physics, University of Washington, Seattle, WA 98195, USA.}

\author{Yu-Ting Chen}
\email{asdasdasdasdasd2578@gmail.com}
\author{Ghanashyam Meher}
\email{ghanashyam@phys.ntu.edu.tw}
\affiliation{Department of Physics and Center for Theoretical Physics,
National Taiwan University, Taipei, Taiwan 106}
\affiliation{Physics Division, National Center for Theoretical Sciences, Taipei 10617, Taiwan}

\author{Berndt M\"{u}ller}
\email{mueller@phy.duke.edu}
\affiliation{Department of Physics, Duke University, Durham, 27708, NC, USA.}
\author{Andreas Sch\"{a}fer}
\email{Andreas.Schaefer@physik.uni-regensburg.de}
\affiliation{Institut f\"{u}r Theoretische Physik, Universit\"{a}t Regensburg, Regensburg, D-93040, Germany.}
\author{Xiaojun Yao}
\email{xjyao@uw.edu}
\affiliation{InQubator for Quantum Simulation (IQuS), Department of Physics, University of Washington, Seattle, WA 98195, USA.}

\preprint{IQuS@UW-21-122}

\date{\today}

\begin{abstract} 
We simulate the local thermalization dynamics for minimally truncated SU(2) pure gauge theory on linear plaquette chains with up to 151 plaquettes using IBM quantum computers. We study the time dependence of the entanglement spectrum, R\'enyi-2 entropy and anti-flatness on small subsystems. The quantum hardware results obtained after error mitigation agree with 
classical simulator results for chains consisting of up to 101 plaquettes. Our results demonstrate the  feasibility of local thermalization studies for chaotic quantum systems, such as nonabelian lattice gauge theories, on current noisy quantum computing platforms.
\end{abstract}

\maketitle

\section{Introduction}
\label{sec:intro}

The Standard Model of particle physics is the underlying theory of electroweak and strong interactions in nature, governing dynamics ranging from the large scale universe to the microscopic scale happening inside a high energy collider. As the symmetry group of the Standard Model is SU(3)$\times$SU(2)$\times$U(1) understanding the dynamical properties of SU($N$) are especially relevant. Perturbation theory and 
Euclidean lattice gauge theory have been very successful in describing perturbative dynamics and equilibrium properties at low baryon density for nonabelian gauge theories, respectively, such that many of the remaining unanswered questions of Standard Model particle physics are related to nonperturbative processes that are far from equilibrium. For recent reviews, see \cite{Bauer:2023qgm,Halimeh:2025vvp}. 

Heavy-ion collisions have provided overwhelming experimental evidence that, already after about 1 fm/$c$, the produced system is well described by a state that appears \XY{locally} thermal in many respects. This remains puzzling, since such rapid thermalization seems difficult to reconcile with several basic considerations, including causality and the unitary time evolution of QCD. Over the past decades, many mechanisms have been proposed to resolve this apparent tension, but no consensus has emerged on the underlying fundamental process. Ref.~\cite{schlichting:2019abc} reviews weak-coupling approaches, while Ref.~\cite{heller:2016gbp} discusses strong-coupling perspectives. An unconventional explanation is presented, for example, in Ref.~\cite{Kharzeev:2026inq}.

It is widely believed that understanding this far-from-equilibrium dynamics of nonabelian gauge theories
requires quantum computing \cite{Davoudi:2025kxb}. Indeed, recent results obtained using digital computers on small lattices suggest that (a) highly excited states of nonabelian gauge fields rapidly thermalize \cite{Yao:2023pht,Ebner:2023ixq,Ebner:2024qtu,Das:2025utp} and (b) that the transition to thermal equilibrium goes through a period of elevated quantum ``magic'', i.e., a quantum magic barrier, which precludes efficient simulation on classical digital computers \cite{Ebner:2025pdm}. The advent of practical quantum computing should very substantially widen the range of problems that can be solved based on first-principle quantum field theory
\cite{Zohar:2012xf, Zohar:2016wmo,Gonzalez-Cuadra:2017lvz,Bender:2018rdp,Abanin:2018yrt, Cloet:2019wre, Halimeh:2020xxe, Rajput:2021trn, Zohar:2021nyc, Bauer:2022hpo, Mildenberger:2022jqr, Mueller:2022xbg, Wang:2022dpp, Belyansky:2023rgh, Beck:2023xhh, DiMeglio:2023nsa, Joshi:2023rvd, Osborne:2023zjn, Ciavarella:2024fzw, Andersen:2024aob, Farrell:2024fit, Farrell:2024mgu, Gonzalez-Cuadra:2024xul, Illa:2024kmf, Davoudi:2024wyv, Spagnoli:2024mib, Ciavarella:2024lsp, Kaufman:2024xnu, Kaufman:2025uas, Aditya:2025yex, Chernyshev:2025lil, Cataldi:2025cyo, DePaciani:2025uzj, Davoudi:2025rdv, Farrell:2025nkx, Froland:2025bqf, Luo:2025qlg, Davoudi:2025rdv, Joshi:2025rha, Joshi:2025pgv, Li:2025sgo, Miranda-Riaza:2025fus, Fontana:2024rux, Santra:2025dsm, Schuhmacher:2025ehh, Xu:2025abo, Yao:2025cxs, Ciavarella:2025bsg, Balaji:2025afl, Huie:2025yzn, Balaji:2025yua, Asaduzzaman:2026jyi, Cao:2026qky, Grieninger:2026bdq, Majcen:2026sfm, Modi:2026syn, Orlando:2026ten, Pato:2026wow, Rouxinol:2026mbh, Spagnoli:2026qni}.

It is therefore of interest to explore the capabilities of currently available quantum computing platforms to reliably simulate the transition of an excited, but initially low-entangled, pure state of the lattice gauge field to a locally thermal state. Thermalization of lattice gauge theories was studied in \cite{Rigol:2007juv, Halimeh:2016fsk, James_2019, deJong:2021wsd, Delacretaz:2021ufg, Mueller:2021gxd, Zhou:2021kdl, Delacretaz:2022ojg, Mueller:2024mmk, Florio:2025xup, Hayata:2026rmv}. However, only a small number of these investigations are based on real-time quantum simulations for gauge theories and were mostly constrained to models without dynamical gauge fields, such as one-dimensional gauge theories (abelian and nonabelian Schwinger model) and the $Z_2$ lattice gauge theory \cite{Mueller:2022xbg, Mueller:2024mmk,Halimeh:2025vvp}. Quantum simulation of nonabelian gauge theory thermalization has only started recently due to the difficulty associated with the nonabelian nature. 

The (2+1)-dimensional Hamiltonian SU(2) lattice gauge theory on a linear plaquette chain represents a chaotic quantum system and satisfies the eigenstate thermalization hypothesis (ETH) at any nonzero lattice coupling constant, as long as the system is large enough \cite{Yao:2023pht,Ebner:2023ixq}. This applies even to the minimally truncated version of the theory, in which the electric field representation on a gauge link is constrained to the values $j=0,\frac{1}{2}$ and the theory can be mapped onto an Ising model with next-to-nearest neighbor transverse field coupling \cite{Yao:2023pht}. The validity of the ETH implies that the expectation values and fluctuations of generic operators with local support thermalize. The thermalization process can be tracked by the time evolution of the entanglement entropy of a local region on the plaquette chain \cite{Ebner:2024mee}.

Another motivation for quantum simulation arises from recent results for the evolution of the entanglement entropy and the anti-flatness of the entanglement spectrum for the SU(2) gauge theory obtained by exact diagonalization of the lattice Hamiltonian, which have revealed a deep connection between the growth rate of the entanglement entropy and the anti-flatness \cite{Ebner:2025pdm}. A large anti-flatness is indicative of a high ``quantumness'' of the transient state of the system. The coincidence of  the anti-flatness maximum with the time of maximal entropy growth indicates the need for full quantum computation during the peak period of thermalization dynamics.

Here we report simulation results on state-of-the-art IBM quantum computers for the local thermalization dynamics of a (2+1)-dimensional SU(2) pure gauge theory on long plaquette chains comprising up to 151 qubits. 
We use the entanglement entropy and the entanglement spectrum as indicators of local thermalization \cite{Mueller:2021gxd}. A variety of error-mitigation tools are used, including dynamical decoupling \cite{Viola:1998gg,Bylander:2011zcm}, Pauli twirling \cite{Geller:2013obn}, and operator decoherence renormalization \cite{Farrell:2023fgd}. Our calculational setup is based on the truncated electric basis of the Kogut-Susskind Hamiltonian \cite{Kogut:1974ag}. 
Let us stress, however, that there exist many different formulations of lattice gauge theories with various dimensions and truncations, some of which could potentially require far fewer resources for quantum computing benchmark tasks than the formulation we use. The exploration of this large field of possibilities is just starting \cite{Chandrasekharan:1996ih, Brower:1997ha, Klco:2019evd, Raychowdhury:2019iki, Buser:2020cvn, Ciavarella:2021nmj, Gonzalez-Cuadra:2022hxt, Kadam:2022ipf, DAndrea:2023qnr, Meth:2023wzd, Muller:2023nnk, Zache:2023cfj, Zache:2023dko, Bergner:2024qjl, Burbano:2024uvn, Carena:2024dzu, De:2024smi, Grabowska:2024emw, Gustafson:2024kym, Halimeh:2024bth, Fontana:2024rux, Cervia:2025vqq, Das:2025utp, Halimeh:2025ivn, Ilcic:2025gel, Illa:2025dou, Jiang:2025ufg, Perez:2025cxl, Siew:2025thj, Yao:2025cxs, Yao:2025uxz, Chen:2026hnh, Siew:2026fax}.

This paper is organized as follows:
In Sec. \ref{sec:H+QC} we specify the Hamiltonian for the truncated SU(2) gauge theory on a linear plaquette chain and explain how to define a subsystem. The quantum circuit for time evolution will be given in Sec.~\ref{sec:method}, as well as the quantum algorithm for subsystem tomography, which will enable us to obtain the entanglement spectrum and calculate antiflatness and various entanglement entropies, such as the R\'enyi-2 entropy. Then in Sec.~\ref{sec:hardware}, we will show the gate counts for one Trotter step in time evolution and explain the techniques for error mitigation and statistical uncertainty estimation. Classical matrix-product-state (MPS)
simulation and quantum hardware results will be presented and discussed in Sec.~\ref{sec:results}. Finally, a summary and conclusions will be given in Sec.~\ref{sec:conclusions}.

\section{SU(2) Hamiltonian on Plaquette Chain and Partitioning}
\label{sec:H+QC}
The Kogut-Susskind Hamiltonian for pure SU(2) gauge theory on a 2-dimensional square lattice is given by 
\cite{Kogut:1974ag}
\begin{align}
    H = \frac{g^2}{2}\sum_{\rm links} E^bE^b - \frac{1}{a^2g^2}\sum_{\rm plaqs} (\square + \square^\dagger) \,,
\end{align}
where $a$ is the lattice spacing, $E^b$ denotes the electric field with SU(2) index $b$ that is implicitly summed and $\square$ represents the plaquette operator defined as the trace of the product of four link operators along the square plaquette. As part of the quantization procedure in the temporal gauge, Gauss's law constraints are imposed on electric fields
\begin{align}
    \sum_{\ell \in v} E_\ell^b \approx 0 \,,
\end{align}
where $\approx$ means constraints to be imposed on states, $\ell \in v$ denotes all links $\ell$ joining the vertex $v$. Physically, Gauss's law means that the state transforms as a singlet under gauge transformation, since the electric field is the generator of gauge transformation \cite{Byrnes:2005qx}. 

In this work, we use the electric field basis in which the electric energy density $\frac{g^2}{2} E^bE^b$ is diagonal. On trivalent lattices such as the plaquette chain considered here, the Hilbert space of singlet states can be uniquely characterized by the irreducible representation (irrep) $j = 0,\frac{1}{2},1,\ldots$ on each link \cite{Kogut:1974ag,Klco:2019evd}. 
Using the results of Ref.~\cite{Klco:2019evd}, the plaquette chain can be mapped onto a one-dimensional spin model with the following Hamiltonian when the local Hilbert space is truncated to the two lowest representations $j\leq j_{\rm max} = \frac{1}{2}$ and local Gauss's law constraints are accounted for at every vertex \cite{Yao:2023pht}, (see similar expressions in Refs.~\cite{Hayata:2021kcp,ARahman:2022tkr} that are equivalent up to a unitary transformation of the basis)
\begin{align}
    H &= 
    J\sum_{i=0}^{N-2} Z_i Z_{i+1} + h_z \sum_{i=0}^{N-1} Z_i +\frac{h_x}{16}\sum_{i=0}^{N-1} X_i 
    \nn\\
    &\quad -\frac{3h_x}{16}\sum_{i=0}^{N-2} [ Z_i X_{i+1} + X_i Z_{i+1}]
    \nn\\
    &\quad +\frac{9h_x}{16}\sum_{i=0}^{N-3}Z_iX_{i+1}Z_{i+2} \,,
    \nn\\
    &\equiv  H_{ZZ} + H_Z + H_X + H_{ZX}+ H_{XZ} + H_{ZXZ} \,,
    \label{eqn:H}
\end{align}
where $J=-\frac{3g^2}{16}$, $h_z = -2J$, $h_x = -\frac{2}{a^2g^2}$. Here, $X$ and $Z$ denote the Pauli-X and -Z matrices, respectively, (similarly for $Y$ later) and $N$ is the number of plaquettes. We have used open boundary conditions for the chain to avoid long connectivity between the first and last qubits in the quantum circuit.\footnote{The boundary conditions are chosen such that the plaquette terms on the first and last plaquettes are of the form $X_0(1-3Z_1)/16$ and $(1-3Z_{N-2})X_{N-1}/16$, respectively, \XY{i.e., setting $Z_{-1}\to 0$ and $Z_{N}\to 0$,} instead of $X_0(1-3Z_1)/4$ and $(1-3Z_{N-2})X_{N-1}/4$ for the boundary conditions of no electric excitation outside the lattice, i.e., $Z_{-1}=Z_{N}=-1$. We choose these particular boundary conditions such that the multi-qubit gates in the quantum circuit are the same in each layer.}
We have defined different Hamiltonian components according to the number and type of Pauli matrices involved, e.g., $H_{ZZ}$ indicates $J\sum_{i=0}^{N-2}Z_iZ_{i+1}$.

In order to study the entanglement entropy and entanglement spectrum, we bipartition the plaquette chain into a subsystem $A$ and its complement. In the spin map, partitioning based on the spin content is natural. In the original SU(2) theory, this partitioning corresponds to cutting the lattice at two vertices that share the same horizontal location. This partitioning also neglects the contributions from different states in the same irrep (e.g., for SU(2) the $m$ states with the same $j$ in $|jm\rangle$ notation) to the entanglement entropy, which are not distillable \cite{VanAcoleyen:2015ccp}.

\section{Quantum Algorithms for Time Evolution and Entanglement Entropy}
\label{sec:method}
\subsection{Trotterized Evolution}
\label{sec:trotter}
We use Trotterization to implement the Hamiltonian time evolution on a quantum computer.
Because the interaction terms on neighboring qubits in the $H_{XZ}$, $H_{ZX}$, and $H_{ZXZ}$ parts of the Hamiltonian do not commute, we further split these terms into smaller components. We also decompose $H_{ZZ}$ because we cannot apply gates for $Z_i Z_{i+1}$ and $Z_{i+1} Z_{i+2}$ simultaneously. Explicit expressions are given by
\begin{align}
    H_{ZZ}^{(n)} &= J \sum_{i=0}^{2i+n+1<N}Z_{2i+n}Z_{2i+n+1} \,, \nn\\
    H_{ZX}^{(n)} &= -\frac{3h_x}{16} \sum_{i=0}^{2i+n+1<N} Z_{2i+n} X_{2i+n+1} \,, \nonumber \\
    H_{XZ}^{(n)} &= -\frac{3h_x}{16} \sum_{i=0}^{2i+n+1<N} X_{2i+n} Z_{2i+n+1} \,, 
\end{align}
for $n=0,\,1$ and
\begin{align}
    H_{ZXZ}^{(m)}
    &= \frac{9h_x}{16}
    \sum_{i = 0}^{3i+m+2<N} Z_{3i+m} X_{3i+m+1} Z_{3i+m+2} \,,
\end{align}
for $m=0,\,1,\,2$.

A single first-order Trotter step for time step $\delta t$ is then given by
\begin{align}
    e^{-iH\delta t} &= U^{(2)}_{ZXZ} U^{(1)}_{ZXZ} U^{(0)}_{ZXZ} U^{(1)}_{XZ} U^{(0)}_{XZ} U^{(1)}_{ZX} U^{(0)}_{ZX} 
    \nn\\
    &\quad \times U_{X} U_{Z} U^{(1)}_{ZZ} U^{(0)}_{ZZ} \,,
    \label{eqn:Trott}
\end{align}
where $U^{(n)}_\alpha = e^{-iH^{(n)}_\alpha\delta t}$. The order of the different unitary operators has been chosen to minimize the two-qubit gate depth.

\begin{figure}[t]
\centering
\includegraphics[width=0.9\linewidth]{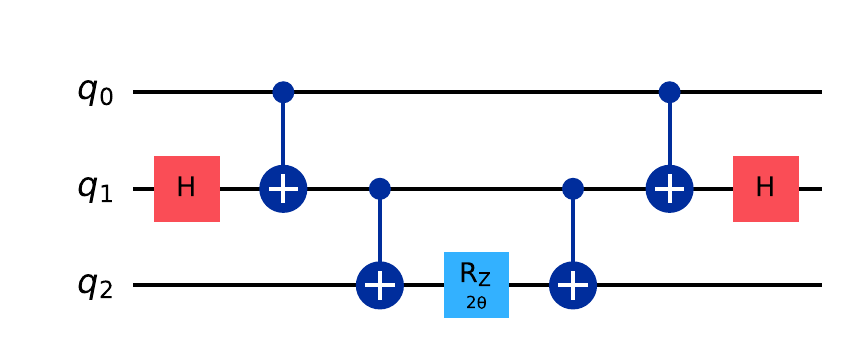}
\caption{Single- and two-qubit quantum gates for the implementation of the three-qubit unitary operator $e^{-i\theta Z_i X_{i+1} Z_{i+2}}$. The $H$ and $R_Z$ in the circuit indicate the standard Hadamard and $Z$-rotation gates and the two-qubit gate is the $\CNOT$ gate with the filled circle as the control and the ``$+$" as the target.
}
\label{fig:time_evol}
\end{figure}

In practical implementations, five
distinct types of unitary evolution operators arise, which are
$$
    e^{-i\theta Z_i} ,\,\, e^{-i\theta X_i},\,\, e^{-i\theta Z_i Z_{i+1}},\,\, e^{-i\theta Z_i X_{i\pm 1}}, \,\, e^{-i\theta Z_i X_{i+1} Z_{i+2}}\,,
$$
where $\theta$ is the product of the relevant coupling constant shown in Eq.~\eqref{eqn:H} and time step $\delta t$. The first four operators can be implemented using standard {\sc Qiskit} gates \cite{Javadi-Abhari:2024kbf}: $R_Z(2\theta,i)$, $R_X(2\theta,i)$, $R_{ZZ}(2\theta,i,i+1)$, and $R_{ZX}(2\theta,i,i\pm 1)$. The last operator, $e^{-i\theta Z_i X_{i+1} Z_{i+2}}$, represents a non-standard three-qubit gate that needs to be constructed from one- and two-qubit gates. We use the well-known construction that is based on the Hadamard, single-qubit $Z$-rotation and controlled-NOT (\CNOT) gates shown in 
Fig.~\ref{fig:time_evol}.

We will start the time evolution with the strong-coupling vacuum state where all the spins of the spin chain are in spin-down configuration, i.e., $|\psi(t=0)\rangle = | \!\downarrow \downarrow \cdots \downarrow \downarrow \rangle$. We note that although this state is highly excited, it has an energy variance that is not much smaller than its mean energy. So some oscillating non-thermal behavior is still expected at late time. We choose this state because it is very easy to prepare on quantum hardware, and thus it allows us to simulate time evolution for a longer time before the hardware noise completely overwhelms the physical signal such that it can no longer be recovered even with error mitigation.\footnote{To make the thermal behavior more manifest at late time without using too many gates in the state preparation, one could use the antiferromagnetic initial state $| \downarrow \uparrow \downarrow \uparrow \cdots \downarrow \uparrow \rangle$.} 

\begin{figure*}[t]
\begin{center}
\includegraphics[width=1\textwidth]{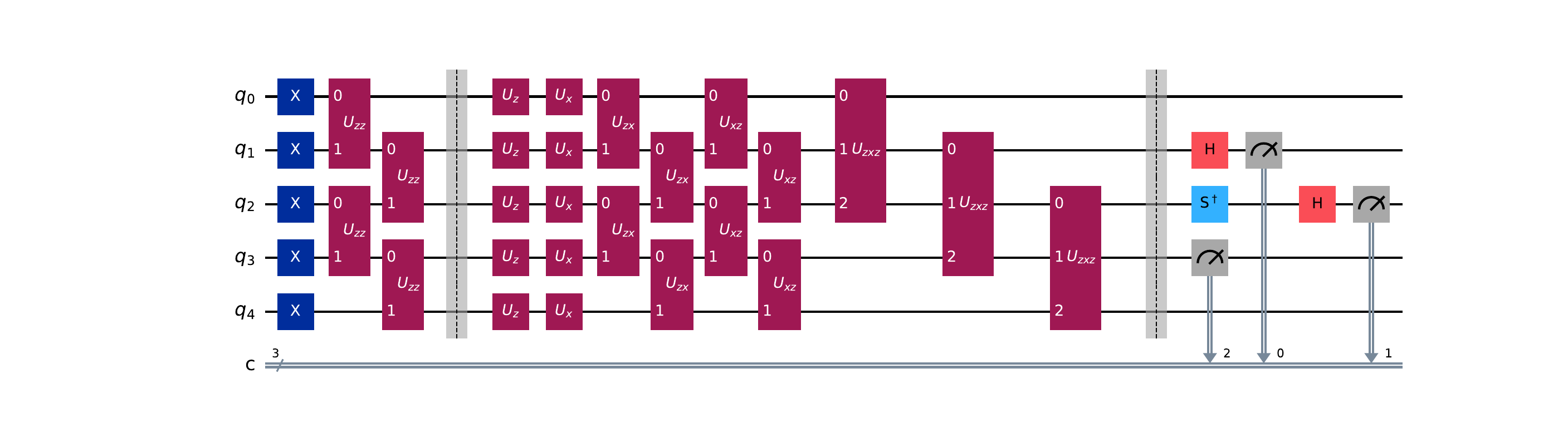} 
\caption{Quantum circuit for the time evolution of an initial strong-coupling vacuum state $|\downarrow \downarrow \downarrow \downarrow \downarrow\rangle$ for a plaquette chain of length $N=5$. $q_0,q_1,q_2,q_3,$ and $q_4$ are the quantum registers and $c$ denotes the classical (measurement) registers.
The single- and two-qubit gates $U_Z=R_Z(2\theta,i)$, and $U_X=R_X(2\theta,i)$, 
$U_{ZZ}=R_{ZZ}(2\theta,i,i+1)$, $U_{ZX}=R_{ZX}(2\theta,i,i+1)$, and $U_{XZ}=R_{ZX}(2\theta,i+1,i)$ correspond to standard {\sc Qiskit} gates. 
The quantum circuit for the three-qubit gate $U_{ZXZ}=e^{-i\theta Z_i X_{i+1} Z_{i+2}}$ is shown in Fig.~\ref{fig:time_evol}. At the end of the time evolution, qubit $q_1$, $q_2$, and $q_3$ representing a $N_A=3$ subsystem are measured in the $X$, $Y$ and $Z$ eigenbasis, respectively, as an example of subsystem state tomography. The vertical gray bars denote barriers, whose effects are discussed in Appendix~\ref{Appendix:barrier}.}
\label{fig:quantcirt}
\end{center}
\end{figure*}

We then evolve the state to an arbitrary time $t$ with $N_t$ Trotter steps and the step size $\delta t = t/N_t$:
\begin{eqnarray}
    |\psi(t)\rangle = \left(\prod_{k=1}^{N_t} e^{-i H \delta t}\right) | \psi(t=0) \rangle \,.
\end{eqnarray}
A quantum circuit for preparing $|\psi(0)\rangle$ and schematically implementing the time evolution of one Trotter step is shown in Fig.~\ref{fig:quantcirt} for a $N=5$ plaquette chain.

\subsection{Subsystem State Tomography}
\label{sec:tomography}
In order to study the entanglement entropy and spectrum, we need to construct the reduced density matrix of a subsystem. We use quantum tomography for this task.
The generic form of the reduced density matrix of a subsystem $A$ consisting of $N_A$ connected spins is given by
\begin{align}
    \rho_{A} &= \frac{1}{2^{N_A}} \sum_{P_A \in \{I,X,Y,Z\}^{\otimes N_A}} \textrm{Tr}[\rho(t) P_A\otimes I_{\bar{A}}]  P_A \,,\nonumber\\
    &= \frac{1}{2^{N_A}} \sum_{P_A \in \{I,X,Y,Z\}^{\otimes {N_A}}} \langle \psi(t)| P_A\otimes I_{\bar{A}} |\psi(t)\rangle  P_A \,,
\end{align}
where the sum runs over all tensor products of $N_A$ Pauli operators including the identity operator and $\bar{A}$ denotes the complement of $A$. To construct $\rho_A$, one needs to measure the expectation values of all Pauli operators at a given time. The computational cost scales exponentially with the subsystem size. As we show later, these measurements can be performed for small subsystems. We expect the exponential scaling not to be the bottleneck as long as one is only interested in local observables.

To measure $\langle \psi(t)| P_A | \psi(t) \rangle $, we rotate each qubit according to the corresponding Pauli operator contained in $P_A$. For $Z$ and $I$, we apply no gates and the measurement is performed directly in the computational basis, which is the eigenbasis of $Z$. For $X$ and $Y$, we rotate the measurement basis before readout as follows:
\begin{eqnarray}
    X \rightarrow Z = H X H\,, \quad
    Y \rightarrow Z = H S^\dagger  Y S H\,,
\end{eqnarray}
where $H$ and $S$ denote the Hadamard and phase gates, respectively. An example of the measurement procedure is schematically shown at the end of the quantum circuit depicted in Fig.~\ref{fig:quantcirt}, where the qubits $q_1$, $q_2$, and $q_3$ are measured in the $X$, $Y$ and $Z$ eigenbases, respectively, to evaluate the matrix element $\langle\psi(t)| X_1Y_2Z_3 |\psi(t) \rangle$.

The results of the measurements are stored as classical bitstrings $\{b_0b_1\ldots b_{N_A-1}\}$ of length $N_A$. The expectation value of the Pauli operator $P_A$ is obtained by averaging over these bitstrings
\begin{eqnarray}
    \langle \psi(t) | P_A |\psi(t) \rangle =\frac{1}{N_\textrm{shots}} \sum_{\{b_i\}} (-1)^{b_0+b_1+\cdots +b_{N_A-1}}\,,
\end{eqnarray}
where $N_\textrm{shots}$ denotes the number of measurements (``shots'') for a given time $t$.

Once the approximate reduced density matrix  $\rho_A$ is reconstructed from the measurement results, the entanglement spectrum can be obtained by diagonalizing $\rho_A$. Furthermore, the R\'enyi-2 entropy 
\begin{eqnarray}
    S_A^{(2)} = - \ln[\textrm{Tr}(\rho_A^2)] \,,
    \label{eq:Renyi-2}
\end{eqnarray}
and the anti-flatness
\begin{eqnarray}
    \mathcal{F}_A = \mathrm{Tr}\!\left(\rho_A^3\right) - \left[\mathrm{Tr}\!\left(\rho_A^2\right)\right]^2 \,,
    \label{eq:Antiflatness}
\end{eqnarray}
can be obtained from the spectrum of $\rho_A$. The anti-flatness serves as a witness of nonlocal magic.
Previous studies have established that magic admits a decomposition into local and genuinely multipartite (nonlocal) components~\cite{Cao:2024nrx, Cao:2023mzo}. Both contributions reflect non-Clifford features of the quantum state. 
The local component can be removed by local unitary transformations, whereas the nonlocal part encodes intrinsically multipartite correlations that cannot be eliminated without destroying entanglement~\cite{Cao:2024nrx, Cao:2023mzo, Qian:2025oit, Andreadakis:2025mfw}. It captures genuinely non-classical correlations beyond both stabilizer structure and entanglement. Consequently, anti-flatness $\mathcal{F}_A$ furnishes a lower bound on the intrinsic classical 
simulation complexity of the system, since $\mathcal{F}_A$ furnishes a lower bound on the nonlocal magic~\cite{Cao:2024nrx}.

\section{Hardware Implementation}
\label{sec:hardware}
\subsection{Gate Counts}
\label{sec:gatecounts}

\begin{figure}[th]
\centering
\subfloat[Two-qubit gate count vs $N$.\label{fig:num_2qubit_gate}]{%
    \includegraphics[width=0.9\linewidth]{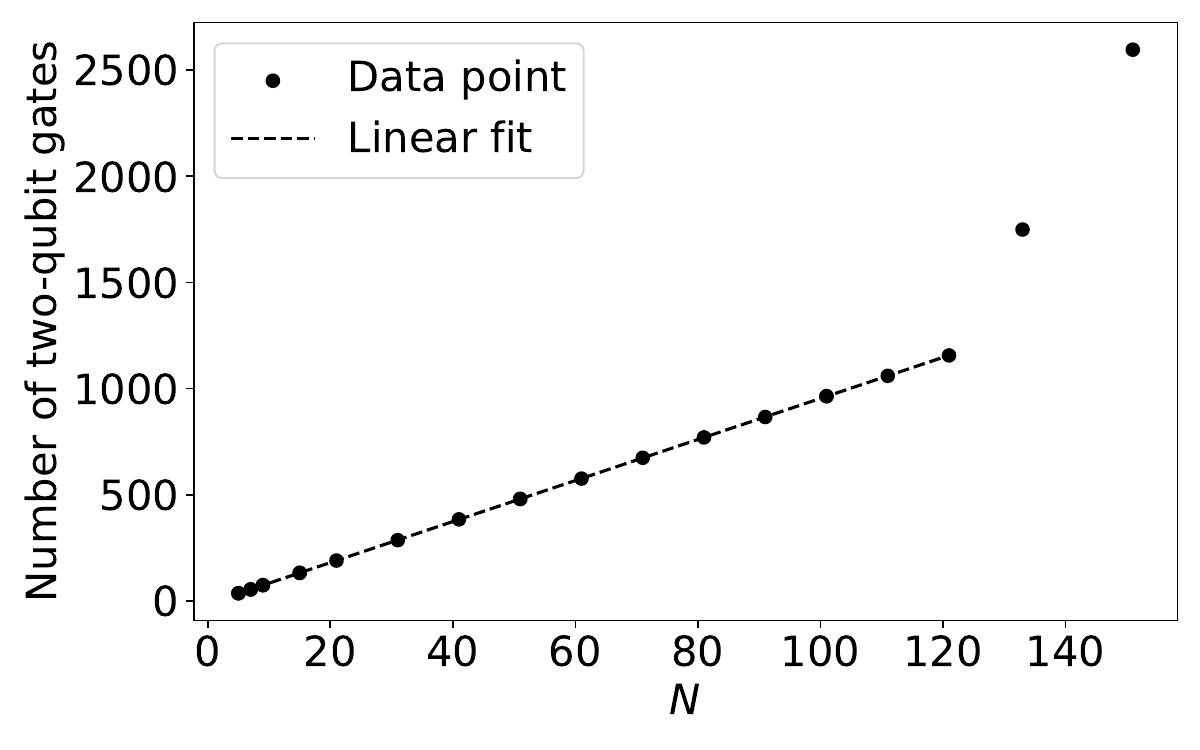}%
}

\subfloat[$N=51$.]{%
  \includegraphics[scale=0.05]{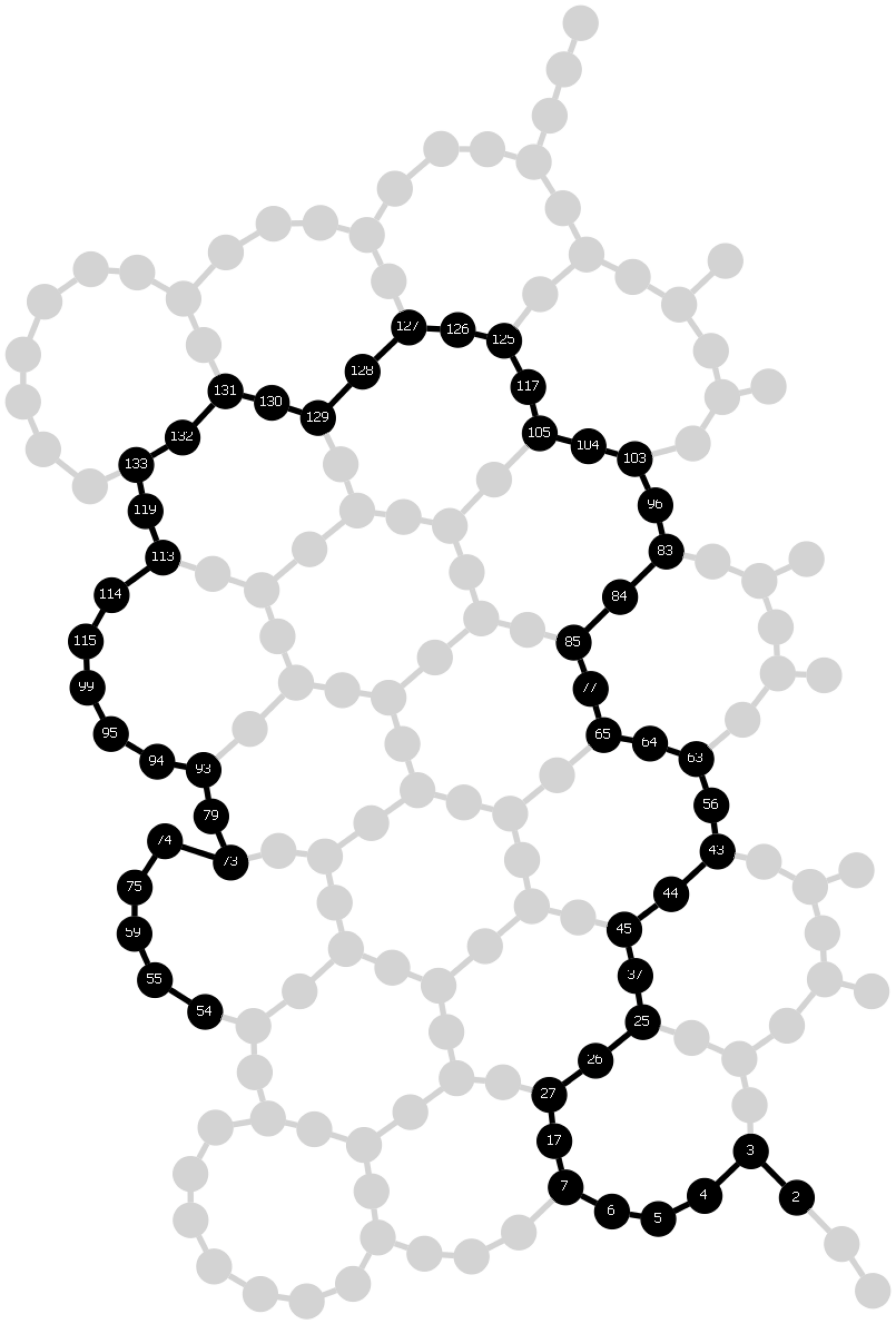}%
} \ \ \ \
\subfloat[$N=133$.]{%
  \includegraphics[scale=0.09]{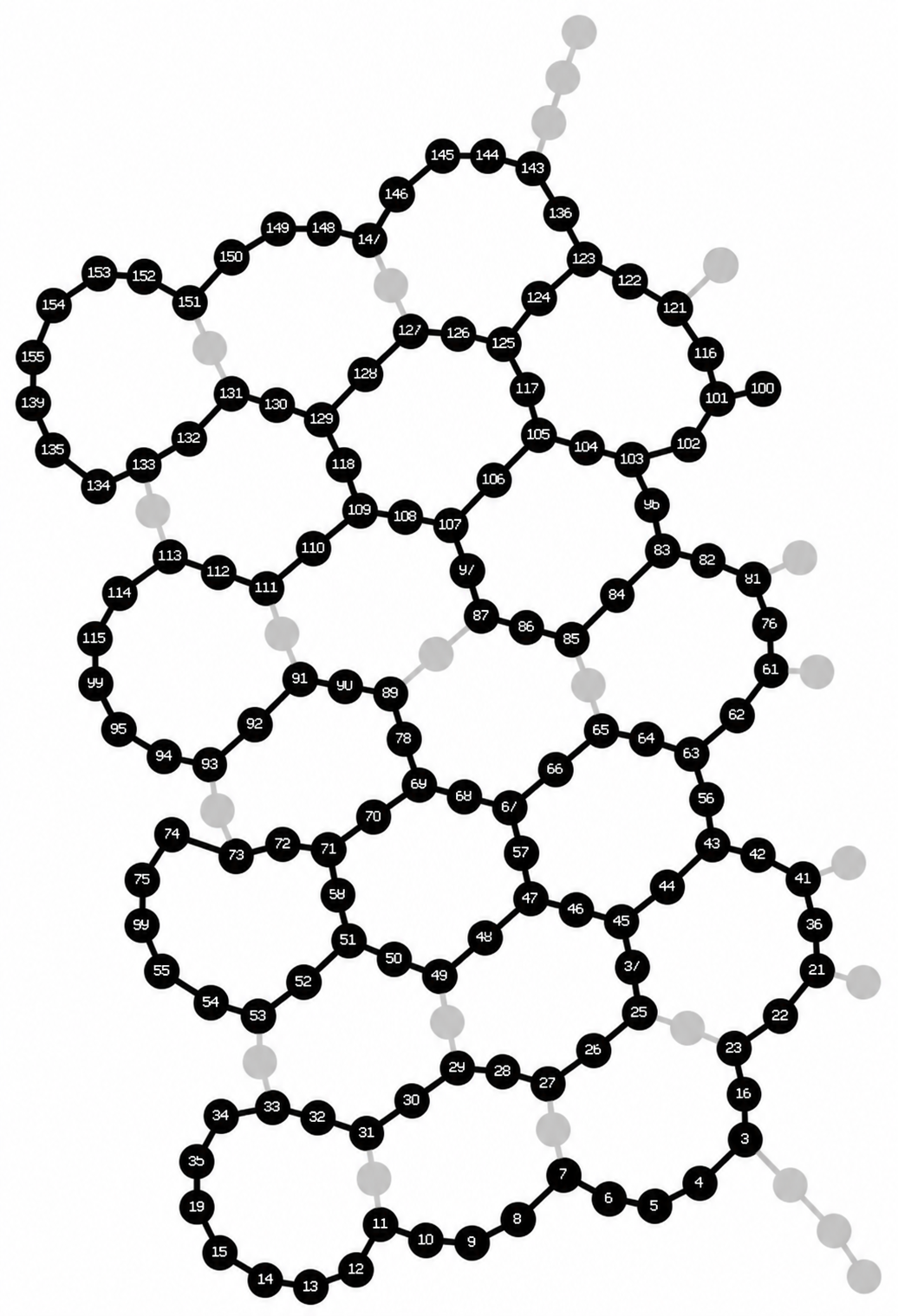}%
}
\caption{(a) Scaling of the two-qubit gate count with system size $N$ for a single Trotter step of the time-evolution quantum circuit. 
The data points (black dots), tabulated in Tab.~\ref{Tab:qubitdepth}, exhibit linear scaling for $N\leq129$, as demonstrated by the linear fit $c_0+c_1N$ (dashed line), with $c_0=-13.082$ and $c_1=9.672$. 
The deviation from linear scaling for $N>129$ occurs because the plaquette chain can no longer be mapped onto a linearly connected sequence of physical qubits on the IBM quantum hardware. 
\JWX{The corresponding qubit mappings are shown for (b) $N=51$ on {\it ibm\_aachen} and (c) $N=133$ on {\it ibm\_boston}. Black (gray) dots denote used (unused) physical qubits}\XY{ that are chosen by the default hardware compiler.}}
\label{fig:2-qubit_gate}
\end{figure}

We simulate dynamics for system sizes of $N =$ 5, 7, 9, 15, 51, 101, 133, and 151 plaquettes (qubits). 
In Tab.~\ref{Tab:qubitdepth}, we summarize the total gate depth, two-qubit gate depth, and the number of two-qubit gates for one Trotter step in the time evolution quantum circuit executed on the IBM quantum computer {\it ibm\_aachen} for various chain lengths. From the table, it is seen that the total gate depth per Trotter step is independent of chain length for $N \leq  101$, with the two-qubit gate depth of 24.
This is attributed to the parallel arrangement of the gates in the quantum circuit as illustrated in Fig.~\ref{fig:quantcirt}, which gives the same number of sequential quantum gate layers for system size $N\le 101$. Due to this parallel arrangement, the number of two-qubit gates required is expected to increase linearly with the system size $N$ for $N\le101$. This behavior is reflected in the third column of Tab.~\ref{Tab:qubitdepth} and is verified in Fig.~\ref{fig:2-qubit_gate}, where the number of two-qubit gates is shown as a function of the system size. The linear increase is demonstrated by a linear fit shown as the dashed line up to $N\le 129$.

\begin{table}[b]
\begin{center}
\scalebox{1}{
\begin{tabular}{|c||c|c|c|}
\hline
\multirow{2}{*}{$N$} & Total & Two-qubit &  Two-qubit \\
& gate depth & gate depth & gate number \\
\hline
\hline
5 &  94    & 24     & 36     \\
\hline
7 &  92    & 24     & 54     \\
\hline  
9 &  95    & 24     & 74     \\ 
\hline
15 &  95    & 24     & 132     \\ 
\hline
51 &  96    & 24     & 480     \\ 
\hline
101 &  96    & 24     & 964     \\ 
\hline
133 &  394    & 149     & 1749     \\ 
\hline
151 &  523    & 198     & 2596     \\ 
\hline
\end{tabular}}
\caption{Total gate depth, two-qubit gate  depth and the number of two-qubit gates per Trotter step for system sizes  $N = 5,\ldots,151$ executed on the IBM quantum computer {\it ibm\_aachen}. \JWX{The abrupt change in the total gate depth at $N=133$ occurs because a linear-chain embedding on the hardware is no longer possible for $N>129$}\XY{, see Fig.~\ref{fig:2-qubit_gate}.}
}
\label{Tab:qubitdepth}   
\end{center}
\end{table}

In contrast, for the system sizes $N=133$ and $151$, Tab.~\ref{Tab:qubitdepth} shows that both the total gate depth and two-qubit gate depth increase sharply. The cause of this rapid increase in gate depth is the breakdown of direct connectivity between the nearest-neighbor qubits on the plaquette chain. 
This connectivity overhead can be understood from the architecture of the {\it ibm\_aachen}, which supports up to 156 qubits. 
The layout of qubits and their connectivity on this quantum chip can be found in Fig. 21 of Ref.~\cite{Mayo:2026xxf}. 
With this layout, one can connect a maximum of 129 qubits linearly as a snake arrangement, which is critical for the parallel implementation in the time evolution quantum circuit shown in Fig.~\ref{fig:quantcirt}.
When the plaquette chain length exceeds $129$, some qubits are not directly connected with their lattice neighbors in the hardware layout, 
As a result, quantum simulation on the hardware needs additional routing operations, which leads to a connectivity overhead and increases the two-qubit gate count and gate depth. This overhead is clearly visible in the deviation of the last two points at $N=133$ and $151$ in Fig.~\ref{fig:2-qubit_gate} from the straight line. It significantly limits the time range over which one can robustly simulate the Hamiltonian evolution, even if error mitigation is applied.

\begin{figure*}[t]
\centering
\includegraphics[scale=0.4]{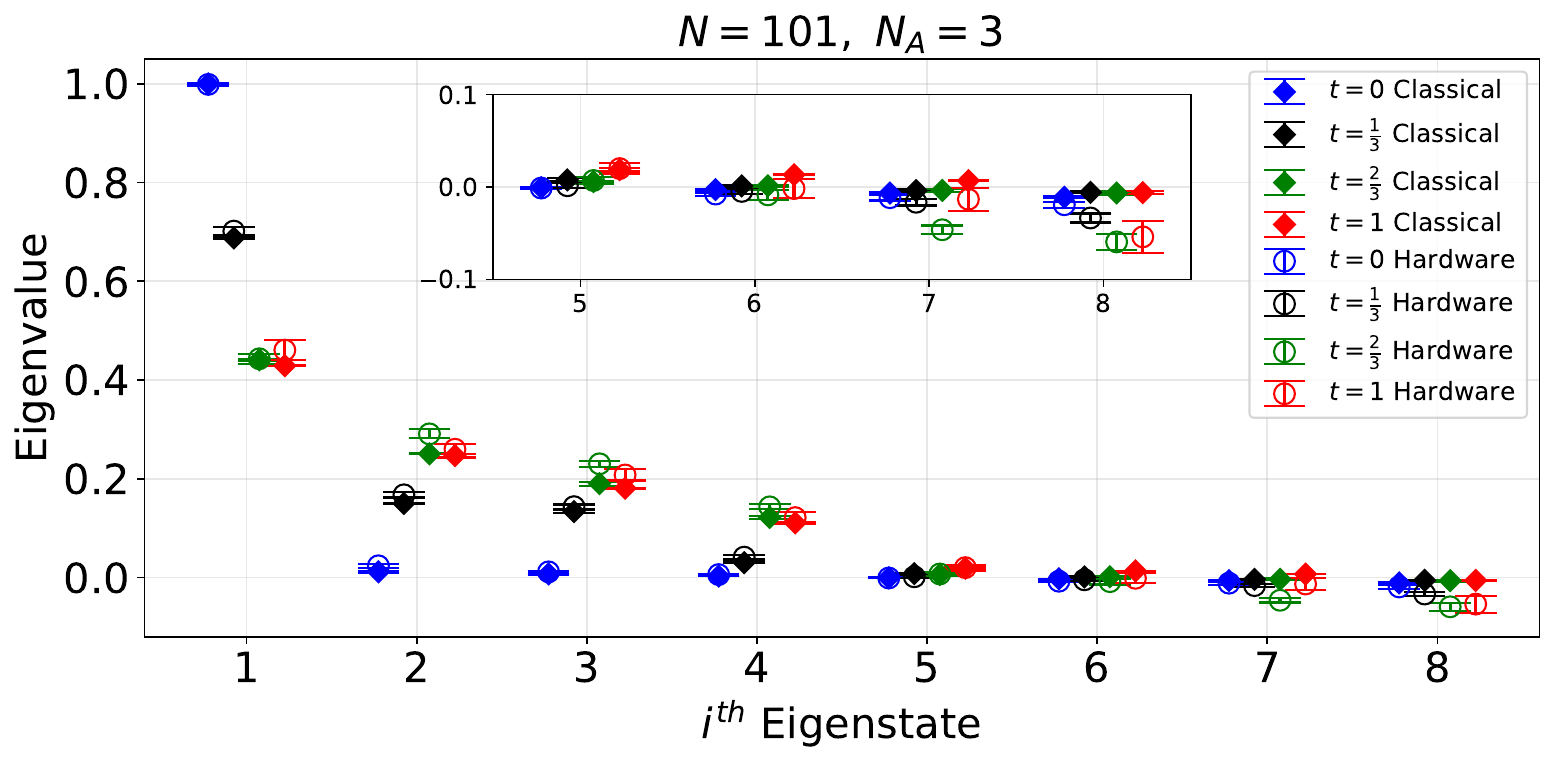} 
\caption{Entanglement spectrum for $N=101$ and $N_A=3$ is shown at time $t= 0, \frac{1}{3}, \frac{2}{3}, 1$. The diamond markers with errorbars denote classical MPS simulator results for $N=101$ \XY{with 20,000 shots}. 
The open circle markers with errorbars represent results obtained from the quantum hardware devices {\it ibm\_boston} and {\it ibm\_torino} by combining measurements of $4,000$ shots from each. The insert figure is a zoom-in of the eigenvalues for the $5^{th}- 8^{th}$ eigenstates. All data points are listed in Tab.~\ref{Tab:E_spectrum} in Appendix~\ref{app:B_data}. 
}
\label{fig:ES_N101NA3}
\end{figure*}

\subsection{Error Mitigation}
\label{sec:error}
Implementation of quantum gates on current quantum hardware introduces noise, contaminating physical signals. When the noise is small, one may still be able to extract the physical signals via error mitigation.
We employ various error mitigation techniques in our quantum calculations such as dynamical decoupling (DD) \cite{Viola:1998gg,Bylander:2011zcm}, 
Pauli twirling (PT) \cite{Geller:2013obn} and operator decoherence renormalization (ODR) \cite{Farrell:2023fgd}.
DD suppresses the decoherence error by sending a sequence of rapid microwave pulses, which keeps the idle qubit 
engaged and the state of the qubit unaltered. PT converts the coherent errors generated by miscalibrated gate rotations, mostly coming from two-qubit gates such as {\CNOT} gates, to incoherent errors.
In PT, noisy {\CNOT} 
gates are sandwiched between randomly chosen Pauli gates such that quantum mechanically they are the same as {\CNOT} gates but error-wise they are different.
The average result of such Pauli rotations 
leads to a normalization prefactor in observables, which can be captured and mitigated by ODR if they are small enough.

Both DD and PT are built-in functions of the {\sc Qiskit} Software Development Kit and can be applied to a circuit by turning on relevant options. For DD, we employ the pulse sequence type ``XY4". For PT, we used 100 different random Pauli-twirled instances of the physical circuit. If the total number of shots is 4,000 then each randomized PT circuit is executed with 40 shots. 
Unlike DD and PT, the ODR error mitigation procedure must be tailored to the specific problem and implemented as post processing. Details of our ODR implementation are provided in Appendix \ref{Appdix:A}.

\subsection{Statistical Uncertainty}
\label{Stat_error}
The error in the measurement of the R\'enyi-2 entropy can be evaluated via standard error propagation as follows:
\begin{align}
    \delta S^{(2)}_A &= \frac{\delta[\textrm{Tr}(\rho_A^2)]}{\textrm{Tr}(\rho_A^2)} \,,
    \label{eq:error_bar} \\
    \delta[\textrm{Tr}(\rho_A^2)] &= \frac{1}{2^{N_A-1}} \bigg[\sum_{P_A} \langle P_A \rangle^2 (\delta P_A)^2 \bigg]^{1/2}\,,
    \label{eq:error_bar_2}
\end{align}
where $\delta P_A$ is the uncertainty in the measurement of $\langle P_A \rangle$. 
$\delta P_A$ is estimated by dividing the measurement data set into several groups (or batches) and computing the variance of the mean values of the batches, from which $\delta P_A$ is determined. We use four batches, which gives us a total of 16 ODR-corrected  reduced density matrices. Details can be found in Appendix~\ref{Appdix:A}.

\begin{figure*}[t]
\centering
\subfloat[$N_A=1$.\label{fig:S2NA1}]{%
  \includegraphics[scale=0.4]{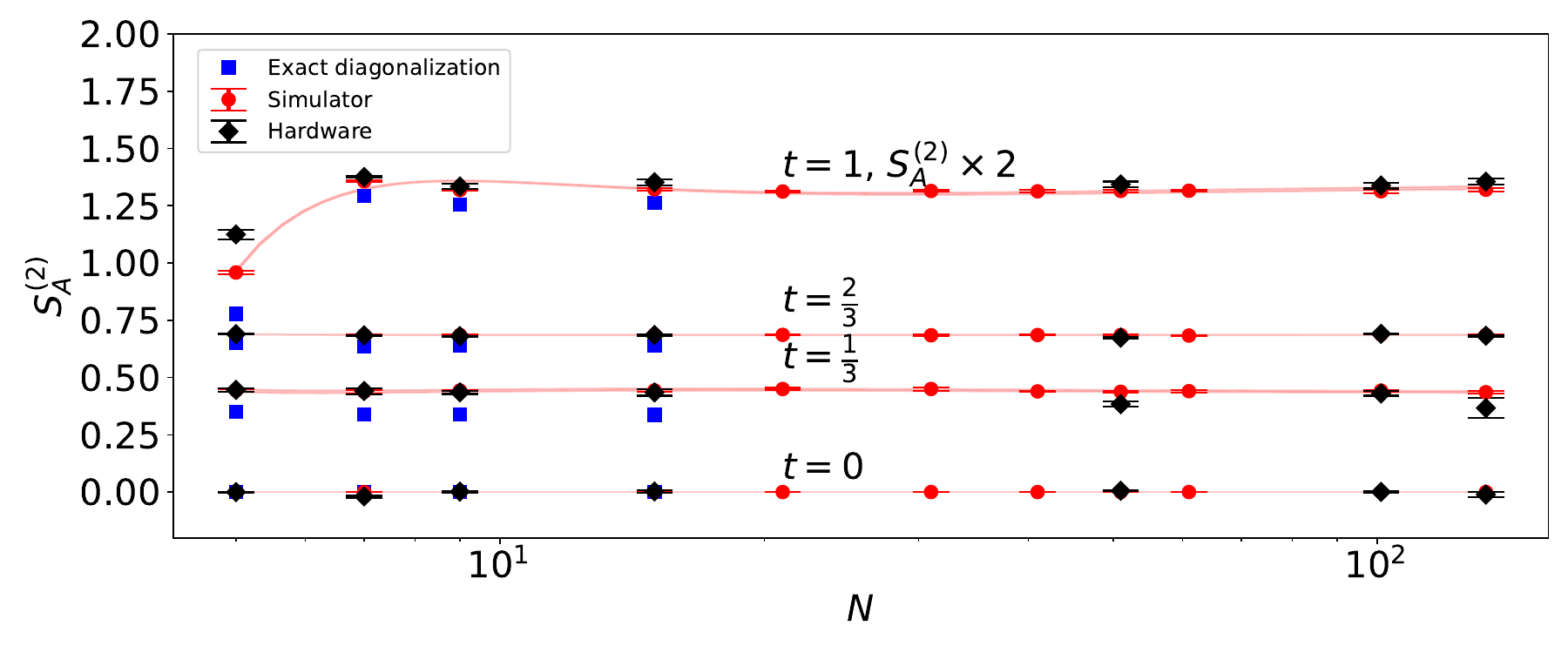}%
}\hfill
\subfloat[$N_A=2$.\label{fig:S2NA2}]{%
  \includegraphics[scale=0.4]{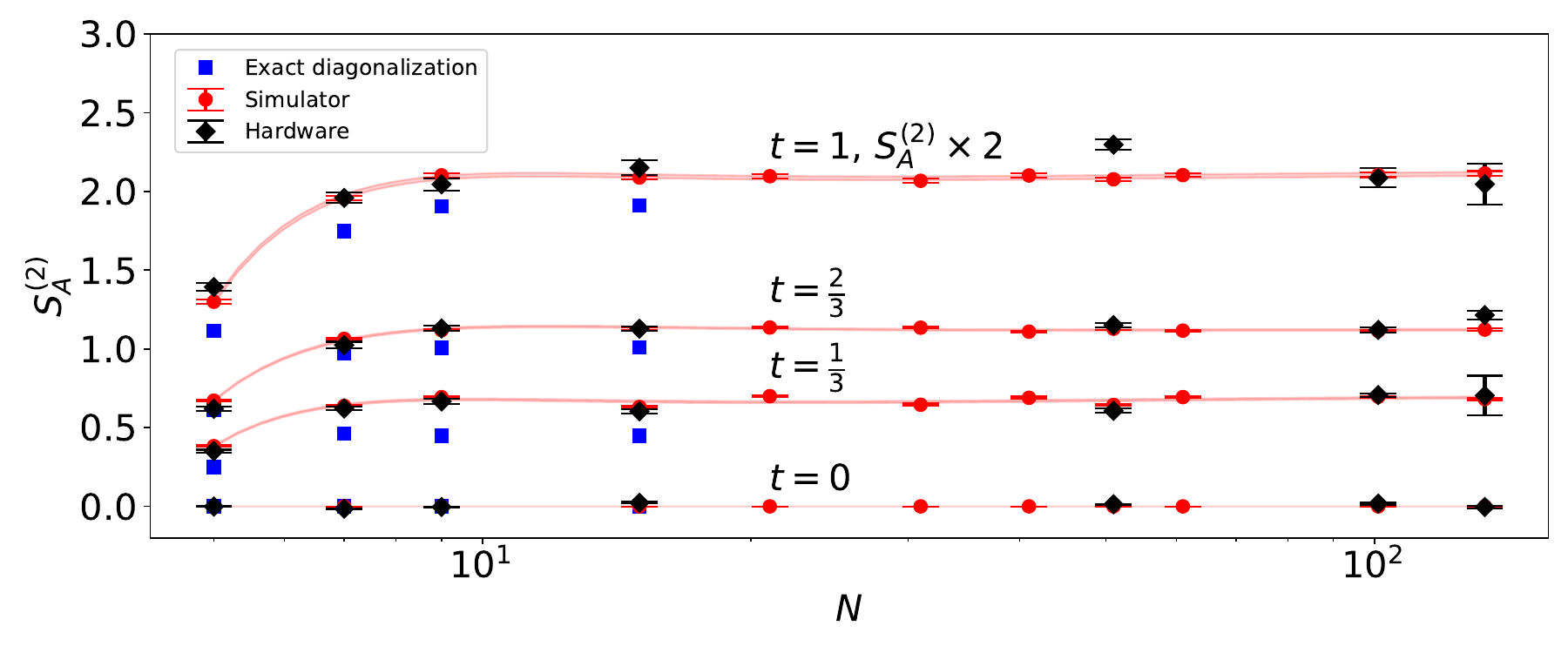}%
}\hfill
\subfloat[$N_A=3$.\label{fig:S2NA3}]{%
  \includegraphics[scale=0.4]{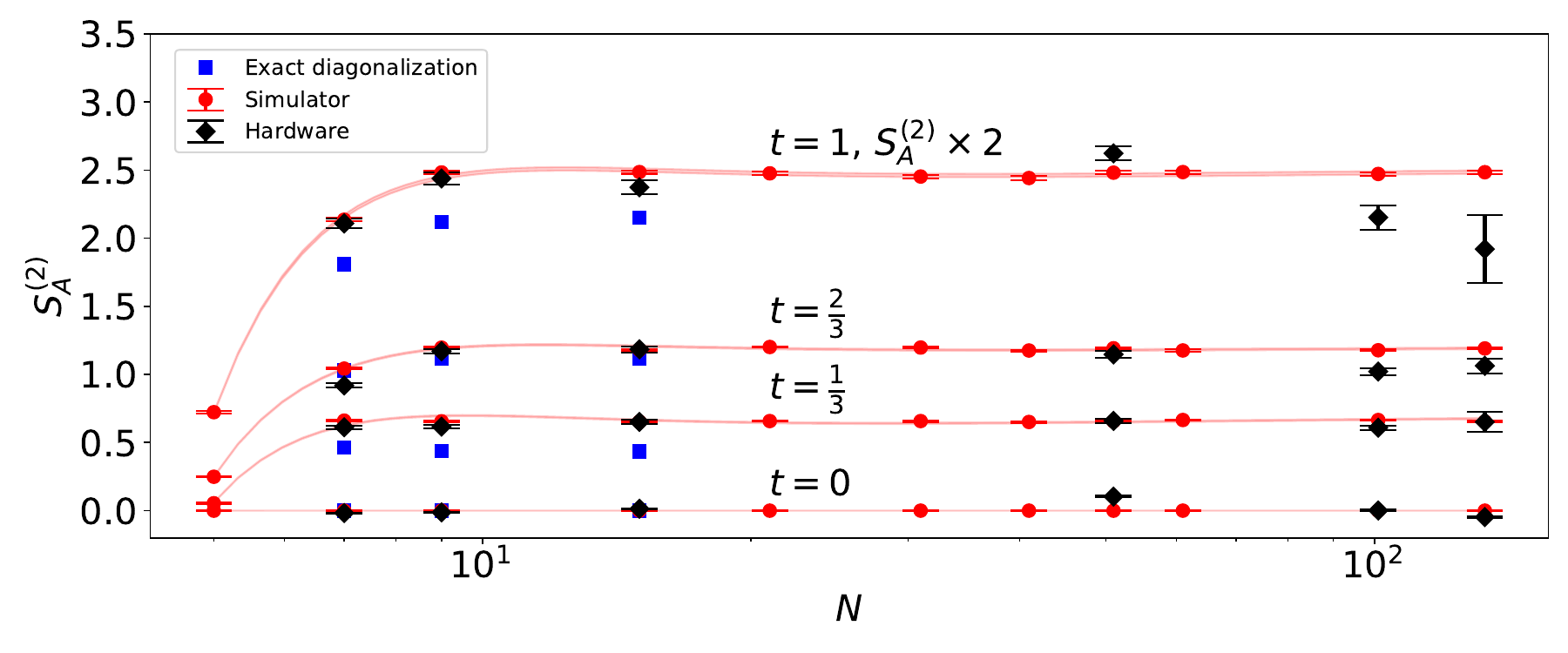}%
}\hfill
\caption{R\'enyi-2 entropy $S_A^{(2)}$ as a function of the system size $N$ at time $t = 0,\,1/3,\, 2/3,$ and 1 for subsystem sizes (a) $N_A=1$, (b) $N_A=2$, and (c) $N_A=3$. Red dots with errorbars denote results obtained from the classical MPS simulator with Trotter and statistical errors, black diamonds with errorbars represent results from real quantum hardware with hardware noise, Trotter and statistical errors, and the blue squares represent exact results obtained from classical exact diagonalization. For clarity, the $t=1$ curves are shown with 
$S_A^{(2)}$ multiplied by a factor of 2.
The light red bands represent the cubic polynomials of $1/N$ fitted from the classical MPS simulator results. 
}
\label{fig:S2vsN}
\end{figure*}

\section{Results}
\label{sec:results}

We time evolve plaquette chains of various lengths as listed in Tab.~\ref{Tab:qubitdepth}, starting from the strong-coupling vacuum state, as explained in Sec.~\ref{sec:trotter}, to the final time $t=1$, which is around the time period featuring rapid entanglement growth as well as large anti-flatness, indicating the need of quantum computing~\cite{Ebner:2025pdm}.
We fix the coupling to be $ag^2 =1.2$, for which the plaquette chain has been shown to exhibit quantum chaos already at length $N\sim20$ \cite{Yao:2023pht}.
We will express every quantity in units of the lattice spacing. The time evolution is performed using a first-order Trotter step size $\delta t =1/3$. 
We construct reduced density matrices for a subsystem $A$ of different sizes $N_A=1, 2,$ and 3, centered on the chain  using state tomography explained in Sec.~\ref{sec:tomography}. We use 8,000 shots per Pauli string measurement (including error mitigation circuits) for $N=101$ and 133 and 4,000 shots for the other $N$s.
By numerically diagonalizing these reduced density matrices, we obtain the entanglement spectra, which further allow us to compute R\'enyi-2 entropy and anti-flatness.

The entanglement spectra obtained from the 
classical MPS simulator with 20,000 shots \XY{per Pauli matrix element evaluation $\langle P_A\rangle$} and from the IBM quantum hardware devices {\it ibm\_torino} and {\it ibm\_boston} (with 4{,}000 shots \XY{per Pauli matrix element evaluation} from each combined) for the $N=101$ plaquette chain and subsystem size $N_A=3$ are shown in Fig.~\ref{fig:ES_N101NA3} and listed in Tab.~\ref{Tab:E_spectrum} in Appendix~\ref{app:B_data}. \XY{The classical results include both Trotterization errors and shot noise.}

For eigenvalues that are not close to zero, the quantum hardware results show reasonable agreement with the 
classical results. Deviations appear at late times for eigenvalues near zero, which we attribute to limited statistics; resolving such small eigenvalues reliably would require a larger number of shots.

The entanglement spectrum only has one nonzero eigenvalue initially since our initial state is a product state. As time evolves, more eigenvalues become nonzero as the dynamics generates entanglement. The entanglement spectra at $t=2/3$ and $1$ are approximately the same, indicating that the system is reaching towards a steady state. The initial state we choose is not near the peak of the energy spectrum, i.e., does not correspond to an infinite temperature state. Thus the entanglement spectrum is not uniform at late time and still contains quite a few very small eigenvalues. If we chose a more highly excited initial state such as the antiferromagnetic state mentioned in Sec.~\ref{sec:trotter}, we would see a more uniformly distributed entanglement spectrum at late time, when the system thermalizes.

Next we compute the R\'enyi-2 entropy $S_A^{(2)}$ as a function of system and subsystem sizes at times $t = 0,\, 1/3,\, 2/3,$ and $1$, as shown in Fig.~\ref{fig:S2vsN}. The top, middle, and bottom panels correspond to subsystem sizes $N_A = 1$, 2, and 3, respectively. For clarity, the $t=1$ results are multiplied by a factor of 2. Red dots with errorbars denote classical MPS simulator results obtained by using 20,000 shots per Pauli matrix element evaluation $\langle P_A\rangle$, black diamonds with errorbars represent results from real quantum hardware, generally using 4,000 shots on {\it ibm\_aachen} or {\it ibm\_fez} or {\it ibm\_kingston} except for $N = 101, 133$ for which we took 8,000 shots in total on {\it ibm\_boston} and {\it ibm\_torino}. Blue squares represent results obtained from classical exact diagonalization that have no errors. Details of uncertainty estimates can be found in Appendix~\ref{Appdix:A}. 
To benchmark quantum hardware results against classical simulator ones \XY{across $N$}, we fit the MPS simulator results at different times $t = 0,\, 1/3,\, 2/3,$ and $1$ using a cubic polynomial function of $1/N$, defined as $f(N) = a + b/N+ c/N^2 + d/N^3$. The fitted values and uncertainties of relevant parameters are listed in Tab.~\ref{Tab:SA_fit} in Appendix~\ref{app:C}. 
The fitted result is consistent with the expectation that the physics of a local subsystem should become independent of the total system size $N$ once $N$ exceeds a certain threshold, because qubits near the spatial boundaries remain outside the causal cone of the subsystem under consideration. Consequently, this expected $N$-independence provides a useful benchmark for assessing the accuracy of the quantum computing results.

The results obtained from real quantum hardware are in good agreement with those from the classical MPS simulator for small subsystem sizes and/or early times for $N\leq 101$, despite the Trotter error indicated by the deviation from the blue squares. As the subsystem size increases, more Pauli string operators need to be  evaluated, many of which have small matrix elements. In order to achieve the same accuracy in evaluating these matrix elements, one has to increase the number of shots. On the other hand, for a fixed subsystem size, we expect more Pauli string matrix elements to become nonzero as time increases. In the beginning, only certain Pauli-Z strings have nonzero matrix elements, since the initial state is a product state in the computational basis. This also leads to demanding more shots at late time. For bigger systems $N=133$ and $151$, the number of CNOT gates and the CNOT depth increase dramatically, as forced by the hardware connectivity and explained in Sec.~\ref{sec:gatecounts} (see also Tab.~\ref{Tab:qubitdepth}). \XY{The accumulated errors becomes too big to be mitigated and} can lead to unphysical results, see Appendix~\ref{app:D_bigN}. 

\begin{figure*}[t]
\centering
\subfloat[$S_A^{(2)}$.\label{fig:N101S}]{%
  \includegraphics[width=0.47\textwidth]{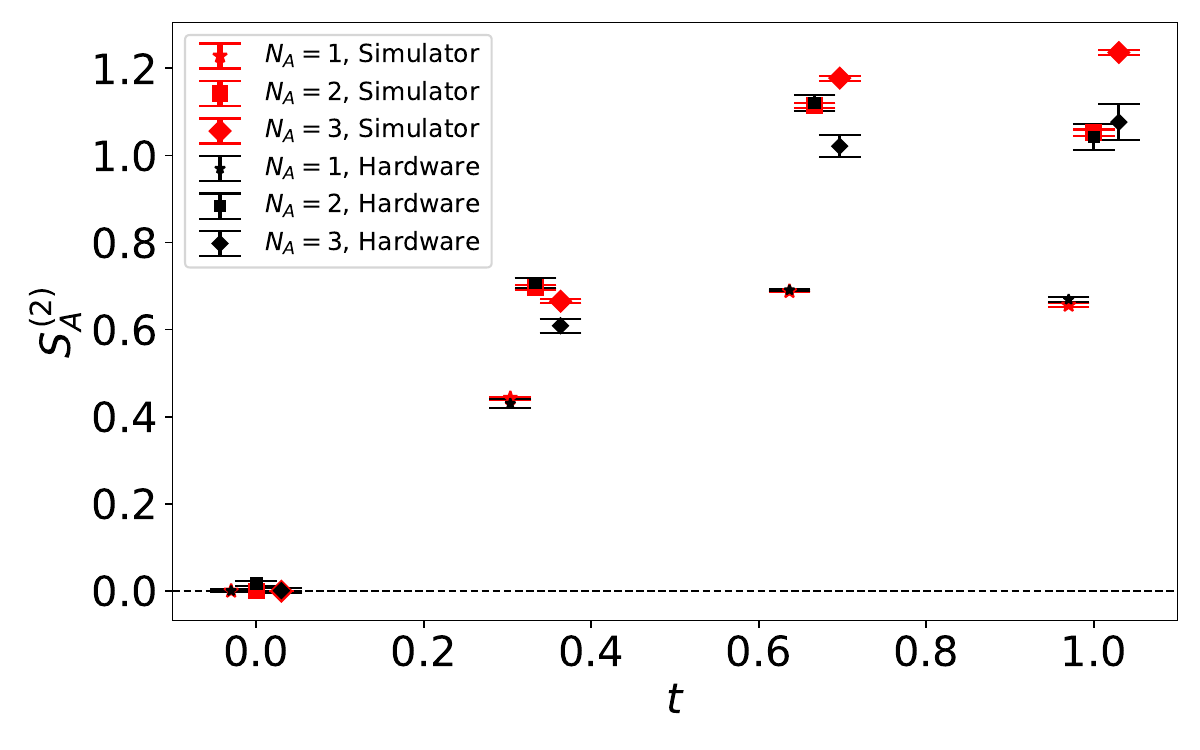} %
}\hfill
\subfloat[$\mathcal{F}_A$.\label{fig:N101F}]{%
  \includegraphics[width=0.48\textwidth]{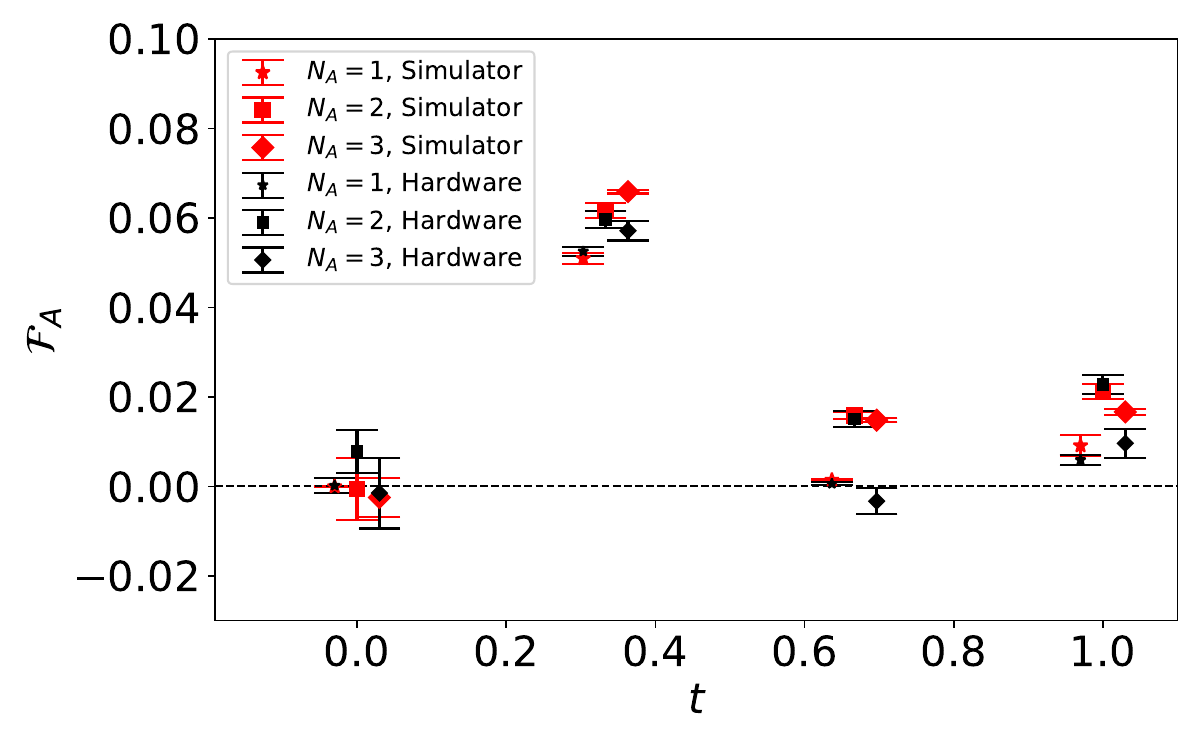} %
}\hfill
\caption{R\'enyi-2 entropy $S_A^{(2)}$ and anti-flatness $\mathcal{F}_A$ as  functions of time for system size 
$N=101$ and subsystem sizes 
$N_A=1,2,$ and 3, shown as stars, squares, and diamonds, respectively. For visual clarity, the data points corresponding to different 
$N_A$ are slightly shifted from 
$t=0,1/3,2/3,$ and 1.
The black markers correspond to the quantum hardware results, whereas the red markers indicate the 
classical simulation results described in Fig.~\ref{fig:S2vsN}. 
}
\label{fig:N101NA123}
\end{figure*}

To more clearly illustrate the time dependence of the entanglement entropy, we plot $S_A^{(2)}$ for $N=101$ at several subsystem sizes and times in Fig.~\ref{fig:N101S}. The R\'enyi-2 entropy approximately reaches a plateau by $t=2/3$, \JWX{consistent with Fig.~\ref{fig:ES_N101NA3}, where the dominant part of the entanglement spectrum also stabilizes around $t\simeq 2/3$. Whether the entanglement spectrum itself follows a Boltzmann distribution \XY{for complete local thermalization} is examined separately in Appendix~\ref{app:Boltzmann}.} The small late-time deviation for $N_A=3$ can be further reduced by increasing the number of measurement shots.

In Fig.~\ref{fig:N101F}, we further show the time dependence of the anti-flatness $\mathcal{F}_A$ for the same system parameters. The anti-flatness develops a pronounced barrier-like peak during the period of rapid growth of the R\'enyi-2 entropy, identifying the regime in which quantum computation may be particularly important for obtaining physically accurate results. This behavior is consistent with that reported in Ref.~\cite{Ebner:2025pdm}, where smaller systems were studied using exact diagonalization.

\section{Conclusions}
\label{sec:conclusions}
In this paper, we performed quantum simulation of local thermalization dynamics of minimally truncated SU(2) pure gauge theory on plaquette chains. We computed the entanglement spectrum, R\'enyi-2 entropy, and anti-flatness as indicators of local thermalization for subsystems consisting of $N_A=1$, $2$, and $3$ plaquettes on chains of lengths up to $N=151$. The quantum simulation was performed on IBM quantum hardware devices {\it ibm\_aachen}, {\it ibm\_fez}, {\it ibm\_boston}, {\it ibm\_torino}, and {\it ibm\_kingston}, subject to machine availability at the time of execution. Time evolution was simulated by using a first-order Trotterization, and reduced density matrices were constructed by state tomography. Various error mitigation techniques were applied, including dynamical decoupling, Pauli twirling and operator decoherence renormalization. For system sizes smaller than $N=129$, we find good agreement between error-mitigated quantum computing results and 
classical MPS simulator results. Beyond $N=129$, accumulated hardware noise exceeds the capability of the implemented error mitigation techniques, and we find large deviations and unphysical behavior in the quantum hardware results. This is attributed to the rapid increase of the number of CNOT gates and CNOT depth as the linear connectivity of the original lattice chain cannot be maintained on the current IBM quantum hardware beyond $N=129$.

In future work, we plan to implement other methods to calculate the R\'enyi-2 entropy and local observables such as randomized measurement tools~\cite{Brydges:2019wut,Elben:2022jvo} and classical shadow analysis~\cite{Huang:2020tih} from quantum hardware measurement results. 
From the perspective of simulating thermalization dynamics, this suffices if one is mostly interested in local thermodynamic observables such as energy density and pressure. We also plan to test other quantum hardware platforms for simulating thermalization dynamics of nonabelian gauge theories.

\begin{acknowledgments}
We thank the IBM Quantum Hub at National Taiwan University for providing computational resources and access to IBM quantum systems used in this work.
 J.W.C., Y.T.C. and G.M. are supported by the National Science and Technology Council of Taiwan under Grant No. 113-2112-M-002-012. B.M. acknowledges support by the U.S. Department of Energy, Office of Science (Grant DE-FG02-05ER41367) and by the National Science Foundation (Project PHY-2434506).
A.S. is supported by the DFG (German Research Foundation, grant 553079183).
X.Y. is supported by the U.S. Department of Energy, Office of Science, Office of Nuclear Physics, InQubator for Quantum Simulation (IQuS) (https://iqus.uw.edu) under Award Number DOE (NP) Award DE-SC0020970 via the program on Quantum Horizons: QIS Research and Innovation for Nuclear Science and acknowledges the discussions at the ``Many-Body Quantum Magic" workshop held at the IQuS hosted by the Institute for Nuclear Theory in Spring 2025. This work was enabled, in part, by the use of advanced computational, storage and networking infrastructure provided by the Hyak supercomputer system at the University of Washington.
\end{acknowledgments}

\appendix
\section{The Effect of Barrier Insertions in the Circuit}
\label{Appendix:barrier}

\begin{table*}[t]
\begin{center}
\footnotesize
\setlength{\tabcolsep}{1pt}
\begin{tabular}{|c||cc|cc|cc|}
\hline \multirow{2}{*}{$N$} 
& \multicolumn{2}{c|}{Total gate depth} 
& \multicolumn{2}{c|}{Two-qubit gate depth} 
& \multicolumn{2}{c|}{Two-qubit gate number} \\
\cline{2-7}
& Before & After & Before & After &  Before & After\\
\hline
\hline
5   & 94  & 90  & 24  & 24  & 36 &  36 \\
\hline
7   & 92  & 88  & 24  & 24  & 54 &  54\\
\hline
9   & 95  & 93  & 24  & 24  & 74 &  74\\
\hline
15  & 95  & 90  & 24  & 24  & 132 & 132 \\
\hline
51  & 96  & 90  & 24  & 24  & 480 &  480\\
\hline
101 & 96  & 91  & 24  & 24  & 964 &  964\\
\hline
133 & 394 & 347 & 149 & 128 & 1749  & 1753\\
\hline
151 & 523 & 510 & 198 & 198 & 2596  & 2657 \\
\hline
\end{tabular}
\caption{
Comparison of the total gate depth and the two-qubit gate depth, defined as the circuit depth obtained after filtering out all non-two-qubit operations, before and after the removal of barriers for circuits of system size $N$ executed on the IBM quantum processor \textit{ibm\_aachen}. The final column gives the corresponding number of two-qubit gates. All numbers are for one Trotter step.}
\label{Tab:depth_comp}
\end{center}
\end{table*}

\XY{In the analysis of gate counts listed Tab.~\ref{Tab:qubitdepth},} we first counted the total number of layers in the directed acyclic graph (DAG) representation of the circuit and then identified those layers that contained at least one two-qubit gate. Under this counting scheme, the insertion of barriers reduced the two-qubit gate depth.

However, since the execution time of single-qubit gates is substantially shorter than that of two-qubit gates, it is common practice to estimate decoherence effects using the depth of the circuit obtained after filtering out all operations other than two-qubit gates. Using this more standard definition, the results are summarized in Tab.~\ref{Tab:depth_comp}. We find that, before and after removing the barriers, the two-qubit gate depths remain unchanged for all system sizes considered, except for the $N=133$ case. In contrast, the total gate depths become slightly smaller after removing the barriers, with reductions of less than $7\%$
 in all cases except $N=133$.

Therefore, removing the barriers is preferable from the perspective of circuit optimization.
Nevertheless, for the circuits studied in this work, the quantitative impact of this modification is marginal.

\section{ODR Error Mitigation and Uncertainty Estimates}
\label{Appdix:A}

The measured expectation values from real quantum hardware are  affected by gate and measurement errors. As a consequence, the measured expectation value deviate from the ideal (noiseless) ones, leading to systematic errors propagating into observables of interest. To mitigate the error, we employ dynamical decoupling \cite{Viola:1998gg,Bylander:2011zcm} and Pauli Twirling \cite{Geller:2013obn} to convert coherent errors to incoherent ones and then use the operator decoherence renormalization (ODR) \cite{Farrell:2023fgd} to mitigate the error during post-processing.
The ODR factor $\eta_O$ for an operator $O$ can be estimated by running a mitigation circuit that has almost the same gate structure and gate count as the physical circuit, but for which the expectation value of the operator is known. By comparing the measured expectation value  $\langle O\rangle_\textrm{meas}^\textrm{mit}$ with the ideal value $\langle O \rangle^{\rm mit}_{\rm ideal}$, both of which are obtained for the mitigation circuit, the ODR factor $\eta_O$ can be determined as follows
\begin{eqnarray}
    \eta_O \langle O\rangle_\textrm{meas}^{\rm mit} = \langle O\rangle_\textrm{ideal}^{\rm mit} \,.
\end{eqnarray}
The error mitigated expectation value of $O$ is then given by
\begin{align}
    \langle O \rangle_{\rm ideal}^{\rm phys} = \eta_O \langle O \rangle_{\rm meas}^{\rm phys} \equiv \frac{\langle O\rangle_\textrm{ideal}^{\rm mit}}{\langle O\rangle_\textrm{meas}^{\rm mit}}\, \langle O \rangle_{\rm meas}^{\rm phys}  \,.
\end{align}

In our calculations, we choose the mitigation circuit to be constructed from half of the time evolution steps applied forward and the other half backward, i.e., 
\begin{equation}
    |\psi^\textrm{mit}(t)\rangle = e^{i Ht/2} e^{-iHt/2}|\psi^\textrm{mit}(t=0)\rangle\,,
\end{equation}
while the physical circuit evolves forward in time in all steps.
The time evolution operators $e^{\pm iHt/2}$ are properly Trotterized as explained in the main text and the gate sequence for $e^{iHt/2}$ is exactly the reverse of that for $e^{-iHt/2}$ such that their product is the identity operator. 

In our practical calculations, the reduced density matrix is constructed at discrete times $t = n\,\delta t$ with step size $\delta t = 1/3$ and $n = 0,1,2,3$. We obtain ODR factors by implementing an equal number of forward and backward time evolution steps with the same size $\delta t$. This provides ODR factors at times $t = 2n\,\delta t = 0,\,2/3,\,4/3,\,2$. 

The ODR factors at $t = 1/3$  and $1$ cannot be obtained directly from the forward--backward calibration procedure because they correspond to an odd number of Trotter steps. They are therefore determined through interpolation. As an example, the ODR factor at $t = 1$ is obtained using the calibrated ODR data at $t = 2/3$ and $t = 4/3$. 
The dataset at each $t$
is divided into four independent batches. Within each dataset, the ODR factors are sorted according to their magnitude, and Python’s built-in piecewise cubic Hermite interpolation is then performed between corresponding ordered 
values. 
Specifically, the largest values from the two datasets are interpolated to obtain the first interpolated ODR factor, the second-largest values are interpolated to obtain the second interpolated factor, and so on. In this way, four interpolated ODR factors are obtained.

This procedure is intentionally conservative. For example, if one were instead to interpolate between the largest value from one dataset and the smallest value from the other, the resulting spread of interpolated ODR factors would be artificially reduced, leading to an underestimate of the uncertainty. By interpolating between corresponding ordered 
values, we preserve the variance structure of the original datasets and obtain a more realistic estimate of the uncertainty.

Nevertheless, the interpolation procedure introduces an additional assumption, namely that the ODR factor varies smoothly with the evolution time. While the use of piecewise cubic Hermite interpolation is designed to preserve monotonicity and avoid spurious oscillations, we acknowledge that this smoothness assumption may introduce an additional systematic uncertainty that is difficult to quantify rigorously.

\begin{figure}[t]
\centering
\subfloat[ODR factors for one-qubit operators.\label{fig:eta_Z}]
{
  \includegraphics[width=0.48\textwidth]{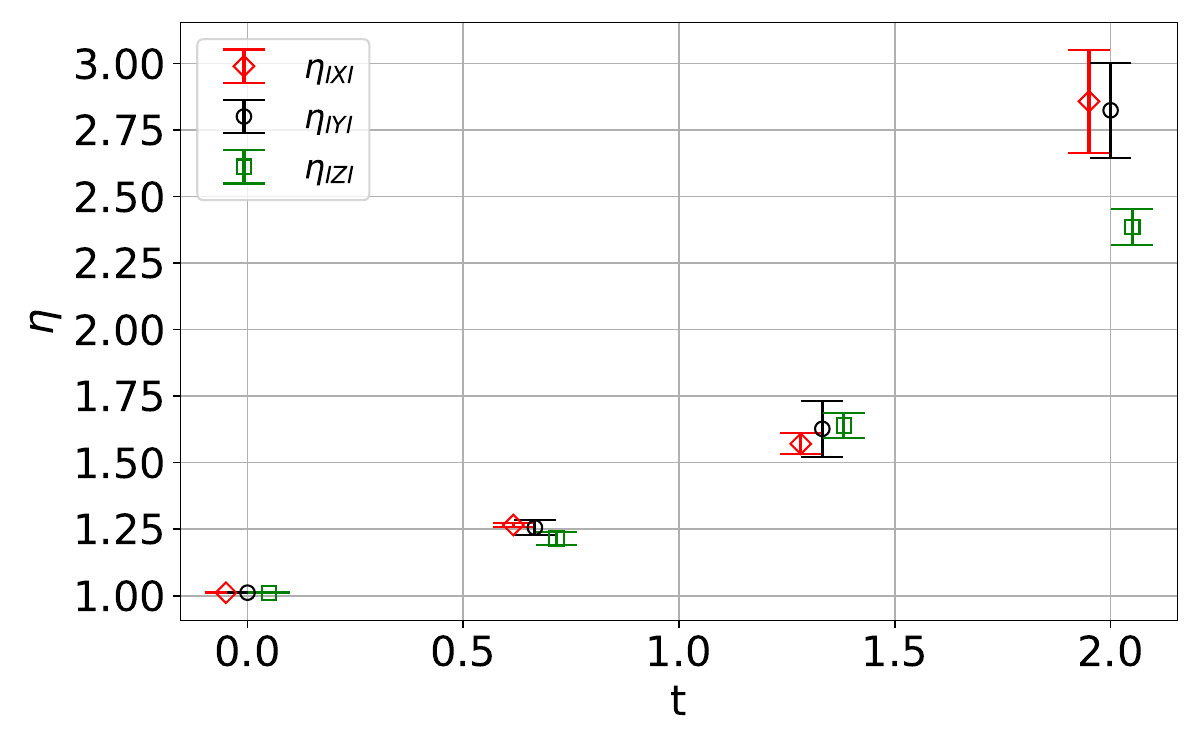} 
}\hfill
\subfloat[ODR factors for three-qubit operators.\label{fig:eta_ZZZ}]{%
  \includegraphics[width=0.48\textwidth]{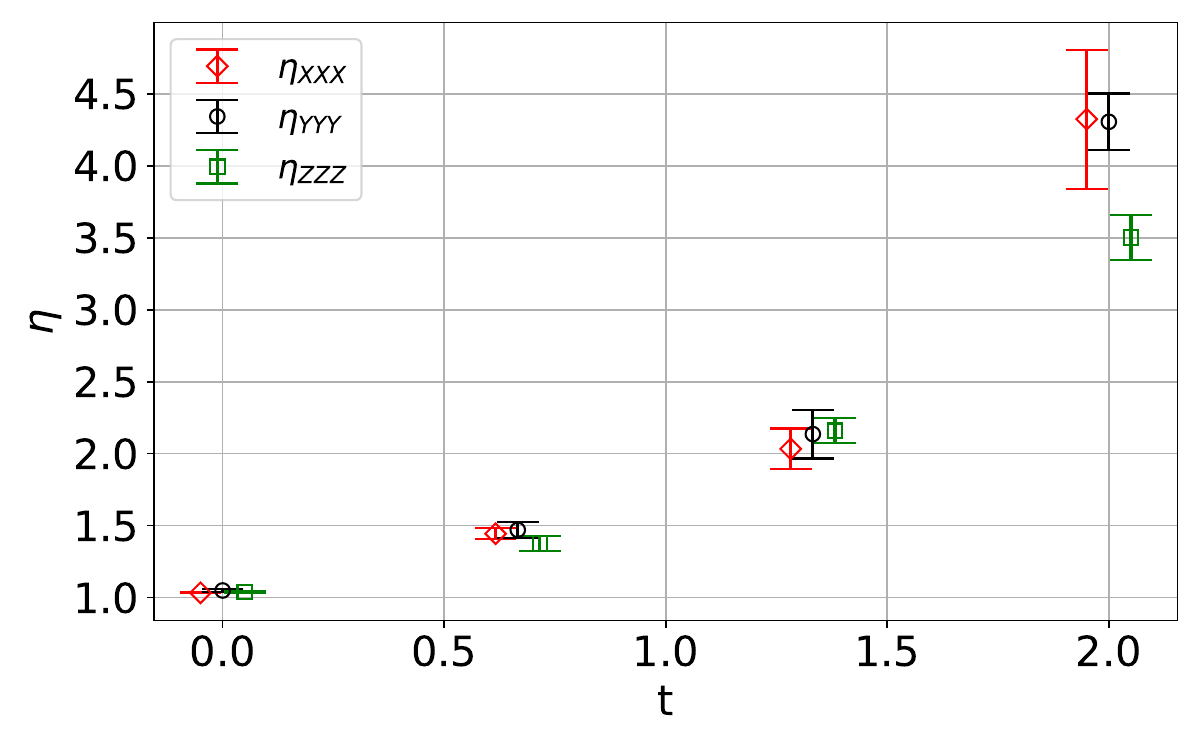} %
}\hfill
\caption{ODR factors for one-qubit and three-qubit operators as a function of total evolved time for the system size $N = 15$, obtained from 12,000 shots on {\it ibm\_boston}. 
The dots denote the ODR factors evaluated at $t = 0,\,2/3,\,4/3,$ and $2$ and are slightly shifted for visibility.
}
\label{fig:ODR}
\end{figure}

In our case the initial state used in the main text is chosen to be the state with all qubits in the $|1\rangle$ state, i.e., $|\!\downarrow \downarrow\cdots \downarrow\downarrow\rangle$ in the spin language, then $\langle O \rangle^{\rm mit}_{\rm ideal}$ vanishes for $O=X$ or $Y$. 
One could apply the $H$ and $HS^\dagger$ gates to the considered qubit before the time evolution and obtain the ODR factors for $\langle X \rangle$ and $\langle Y \rangle$, respectively. 
To very good precision, the ODR factors for $\langle X \rangle$ and $\langle Y \rangle$ for  $t\le 1$
are the same as that for $\langle Z \rangle$, as shown in Fig.~\ref{fig:eta_Z}. In a practical error mitigation procedure, we use the estimated ODR factor of $Z$ for the ODR factors of $X$ and $Y$. 

By the same reasoning, which is supported by Fig.~\ref{fig:eta_ZZZ}, ODR factors for multi-qubit Pauli strings containing $X$ and/or $Y$ operators are inferred from those of Pauli strings in which the corresponding $X$ and/or $Y$ operators are replaced by $Z$. Consequently, for a subsystem size $N_A$, it suffices to determine the ODR factors from \JW{
$4^{N_A}-1$
} Pauli strings only consisting of tensor products of $I_i$ and $Z_i$ ($i\in A$) with at least one $Z$ factor.

Instead of estimating the statistical uncertainties of the ODR factor and the Pauli matrix element separately and then combining them, we divide the data of bitstrings into four batches. Each batch yields one ODR factor value and one Pauli matrix element value, resulting in a total of $4 \times 4 = 16$ ODR-corrected matrix element values. The spread of these 16 values is used to estimate the statistical uncertainty.

The procedure for combining the four batch results to determine the uncertainty of the ODR-corrected reduced density matrix is subtle. If the ODR factors were exactly unity and we had $N_b$ batch means, the uncertainty of a measured matrix element would simply be the standard deviation of these means divided by $\sqrt{N_b}$.
In the present case, however, the ODR factors deviate from unity and are themselves obtained from $N_b$ batches. One can therefore construct $N_b^2$ ODR-corrected matrix elements by combining the estimated matrix element from each batch with the estimated ODR factor from each batch. The uncertainty \JWX{(i.e., the standard error of the mean)} of the corrected matrix element is then obtained by dividing the standard deviation of these $N_b^2$ values by $\sqrt{N_b}$, rather than by $N_b$.
This scaling arises because, if one instead propagated uncertainties by first computing the errors of the matrix elements and the ODR factors separately and then combining them in quadrature, each contribution would scale as $1/\sqrt{N_b}$. Constructing the $N_b^2$ corrected samples is simply a convenient way to implement this error propagation. This prescription of the uncertainty estimate also correctly reproduces the special case in which the ODR factors are exactly unity.

For $N=101$ and $133$, an initial set of $4{,}000$ shots was taken, followed by an additional $4{,}000$ shots, either on a different quantum device or on the same device at a significantly later time. Since the uncertainties associated with these two datasets may differ, we treat them separately. The first $4{,}000$ shots are divided into four batches, yielding $16$ ODR-corrected matrix elements; the same procedure is applied to the second $4{,}000$ shots. We then combine all $32$ corrected matrix elements, compute their standard deviation, and divide by $\sqrt{8}$ to obtain the final uncertainty.    


\section{Entanglement Spectrum Data}
\label{app:B_data}
The data points for the entanglement spectrum shown in Fig.~\ref{fig:ES_N101NA3} are listed in Tab.~\ref{Tab:E_spectrum}.

\begin{table}[t]
\centering
\begin{tabular}{|c|c|c|}
\hline
\multicolumn{3}{|c|}{$t=0$} \\
\hline
$i^{th}$ Eigenvalue & Classical & Hardware \\
\hline
1 & $1.000 \pm 0.000$ & $0.999 \pm 0.002$ \\
2 & $0.011 \pm 0.004$ & $0.024 \pm 0.003$ \\
3 & $0.007 \pm 0.002$ & $0.011 \pm 0.003$ \\
4 & $0.003 \pm 0.001$ & $0.006 \pm 0.002$ \\
5 & $-0.001 \pm 0.002$ & $-0.001 \pm 0.001$ \\
6 & $-0.003 \pm 0.001$ & $-0.008 \pm 0.002$ \\
7 & $-0.007 \pm 0.002$ & $-0.012 \pm 0.003$ \\
8 & $-0.011 \pm 0.002$ & $-0.019 \pm 0.005$ \\
\hline
\multicolumn{3}{|c|}{$t=\frac{1}{3}$} \\
\hline
$i^{th}$ Eigenvalue & Classical & Hardware \\
\hline
1 & $0.688 \pm 0.004$ & $0.702 \pm 0.011$ \\
2 & $0.150 \pm 0.003$ & $0.167 \pm 0.070$ \\
3 & $0.133 \pm 0.007$ & $0.143 \pm 0.004$ \\
4 & $0.030 \pm 0.003$ & $0.041 \pm 0.006$ \\
5 & $0.008 \pm 0.003$ & $0.001 \pm 0.004$ \\
6 & $0.001 \pm 0.003$ & $-0.005 \pm 0.004$ \\
7 & $-0.004 \pm 0.002$ & $-0.017 \pm 0.003$ \\
8 & $-0.006 \pm 0.002$ & $-0.034 \pm 0.070$ \\
\hline
\multicolumn{3}{|c|}{$t=\frac{2}{3}$} \\
\hline
$i^{th}$ Eigenvalue & Classical & Hardware \\
\hline
1 & $0.441 \pm 0.005$ & $0.443 \pm 0.010$ \\
2 & $0.250 \pm 0.002$ & $0.291 \pm 0.009$ \\
3 & $0.190 \pm 0.008$ & $0.230 \pm 0.012$ \\
4 & $0.122 \pm 0.005$ & $0.143 \pm 0.005$ \\
5 & $0.006 \pm 0.001$ & $0.007 \pm 0.004$ \\
6 & $0.001 \pm 0.002$ & $-0.009 \pm 0.003$ \\
7 & $-0.004 \pm 0.003$ & $-0.046 \pm 0.007$ \\
8 & $-0.006 \pm 0.003$ & $-0.059 \pm 0.010$ \\
\hline
\multicolumn{3}{|c|}{$t=1$} \\
\hline
$i^{th}$ Eigenvalue & Classical & Hardware \\
\hline
1 & $0.429 \pm 0.003$ & $0.461 \pm 0.016$ \\
2 & $0.247 \pm 0.008$ & $0.260 \pm 0.004$ \\
3 & $0.181 \pm 0.002$ & $0.208 \pm 0.010$ \\
4 & $0.111 \pm 0.002$ & $0.121 \pm 0.016$ \\
5 & $0.019 \pm 0.004$ & $0.020 \pm 0.009$ \\
6 & $0.013 \pm 0.001$ & $-0.002 \pm 0.010$ \\
7 & $0.007 \pm 0.001$ & $-0.013 \pm 0.070$ \\
8 & $-0.006 \pm 0.003$ & $-0.054 \pm 0.015$ \\
\hline
\end{tabular}
\caption{Entanglement spectra for $N=101$ and $N_A=3$ at different times. The classical results at $N=101$ are obtained 
\XY{
from MPS simulator with 20,000 shots and} are compared with data obtained from the quantum hardware devices {\it ibm\_boston} and {\it ibm\_torino}, combining 4,000 shots from each.}
\label{Tab:E_spectrum}
\end{table}

\section{Fitting Classical Simulator Results}
\label{app:C}
The parameter values fitted to the classical MPS simulator results 
for the R\'enyi-2 entropy shown in Fig.~\ref{fig:S2vsN} are listed in Tab.~\ref{Tab:SA_fit}. Only central values are shown in these tables.

\begin{table}[ht]
\centering
\begin{tabular}{|c| c|| c| c| c| c|}
\hline
$N_A$ & $t$ & $a$ & $b$ & $c$ & $d$ \\
\hline

3 & 0   & -0.000  & 0.005   & -0.061  & 0.192  \\
3 & $\frac{1}{3}$ & 0.700   & -3.803  & 72.464  & -347.196  \\
3 & $\frac{2}{3}$ & 1.206 & -2.345 & 59.715 & -359.470 \\
3 & 1   & 1.256  & -2.112 & 51.478 & -316.673 \\

\hline

2 & 0   &-0.000 &0.001   &-0.011  &0.031  \\
2 & $\frac{1}{3}$ &0.707 &-2.398  &38.308  &-172.269  \\
2 & $\frac{2}{3}$ &1.126 &-0.801  &25.615  &-164.763  \\
2 & 1   &1.066 &-1.557  &30.734  &-166.842  \\

\hline

1 & 0   &-0.000  &-0.003  &0.038 &-0.140 \\
1 & $\frac{1}{3}$ &0.433  &0.552  &-6.462 &19.823 \\
1 & $\frac{2}{3}$ &0.686  &0.014  &-0.278 &1.507 \\
1 & 1   &0.674  &-1.446 &27.236 &-123.975 \\

\hline
\end{tabular}
\caption{\XY{Obtained parameter values for the function $f(N) = a + b/N+ c/N^2 + d/N^3$ that fits the $N$-dependence of the} R\'enyi-2 entropies for different subsystem sizes $N_A$ and times $t$ obtained from classical MPS simulations in Fig.~\ref{fig:S2vsN}. Only central values of the fitted parameters are shown.}
\label{Tab:SA_fit}
\end{table}

\section{Results for 133 and 151 qubits}
\label{app:D_bigN}

\begin{figure*}[t]
\subfloat[$S_A^{(2)}$.\label{fig:N133S}]{%
  \includegraphics[width=0.47\textwidth]{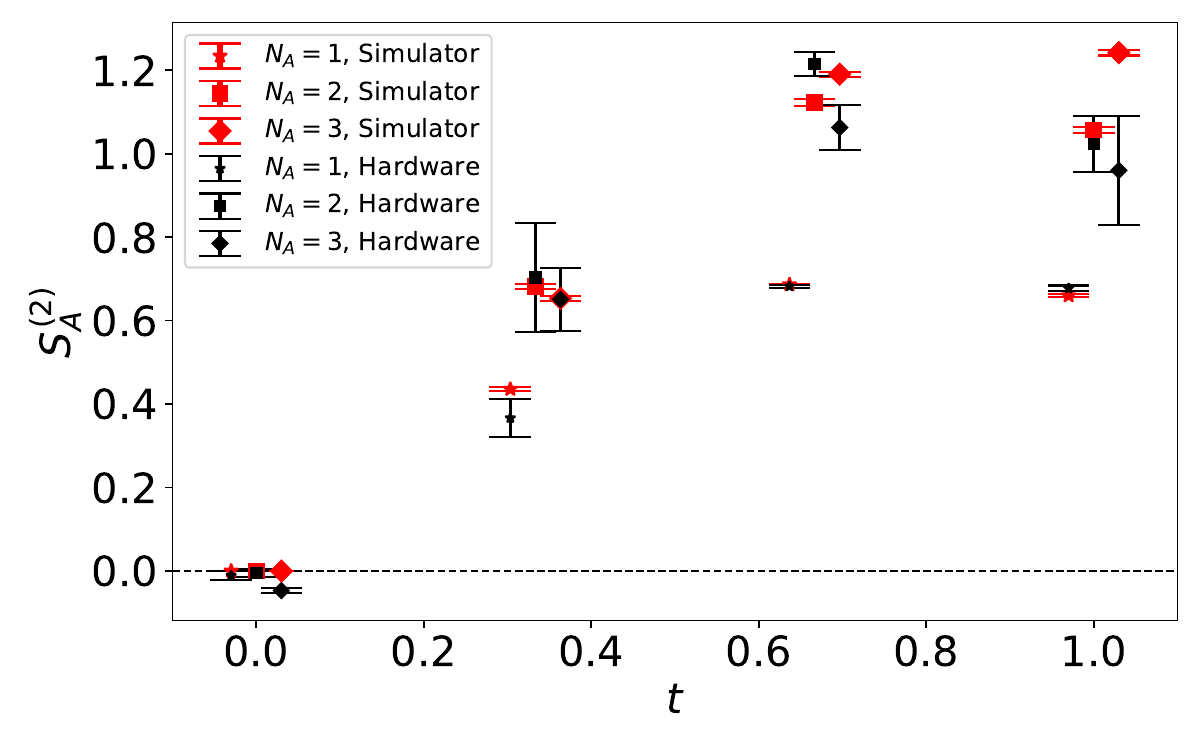} %
}\hfill
\subfloat[$\mathcal{F}_A$.\label{fig:N133F}]{%
  \includegraphics[width=0.48\textwidth]{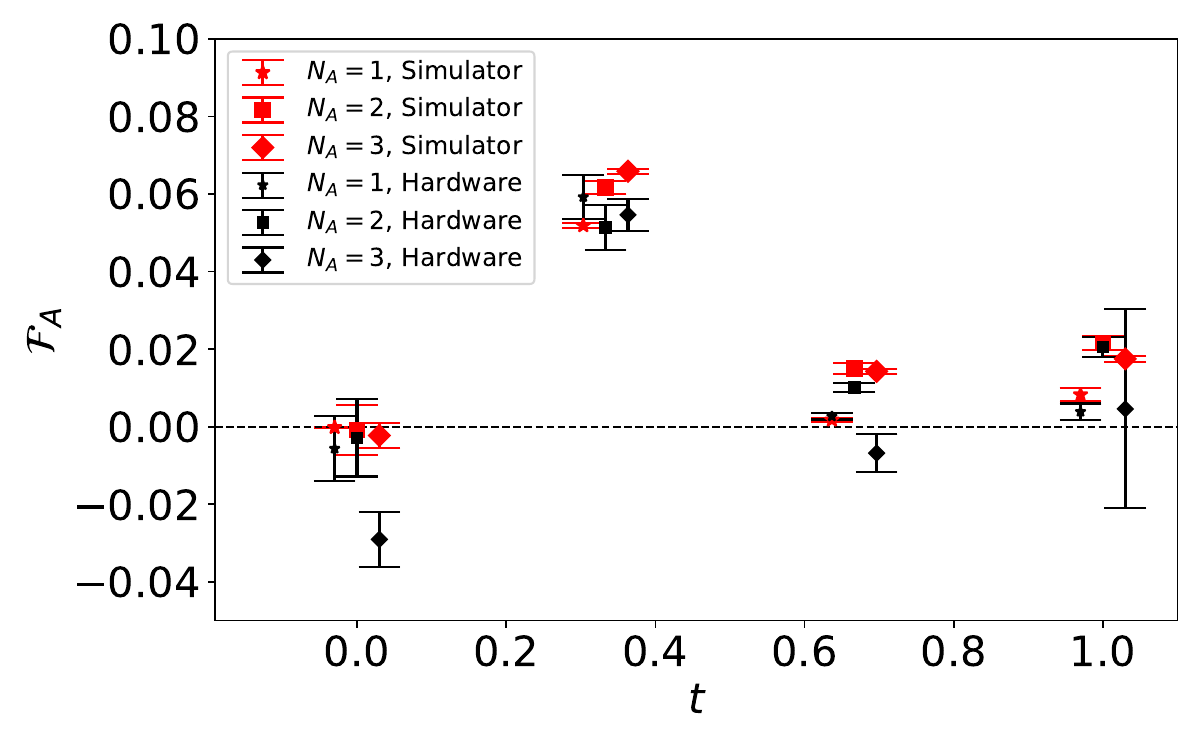}  %
}\hfill
\caption{R\'enyi-2 entropy and anti-flatness of entanglement spectrum as functions of time for system size $N=133$ and subsystem sizes $N_A=1$, $2$, and $3$. The notation follows that of the 
$N=101$ case shown in Fig.~\ref{fig:N101NA123}, except that the hardware experiments were carried out on the IBM quantum hardware ``{\it ibm\_boston}" using 8,000 shots.
}
\label{fig:N133NA123}
\end{figure*}

Adopting the same notation as in Fig.~\ref{fig:N101NA123}, the time dependence of the R\'enyi-2 entropy $S_A^{(2)}$ and the anti-flatness $\mathcal{F}_A$ for system size $N = 133$ is presented in Fig.~\ref{fig:N133NA123}. The $N=133$ data remain consistent with the 
classical MPS simulation results within the (much larger) errorbars.

\begin{figure*}[t]
\subfloat[$S_A^{(2)}$.\label{fig:N151S}]{%
  \includegraphics[width=0.47\textwidth]{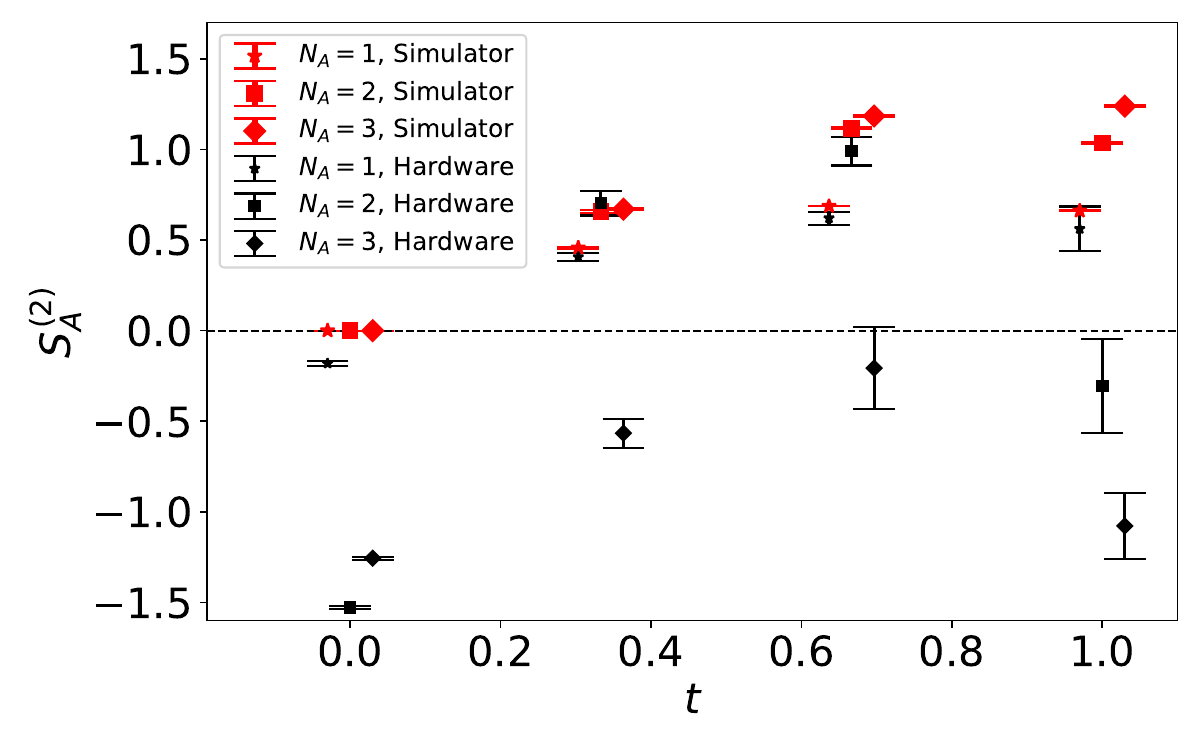} %
}\hfill
\subfloat[$\mathcal{F}_A$.\label{fig:N151F}]{%
  \includegraphics[width=0.48\textwidth]{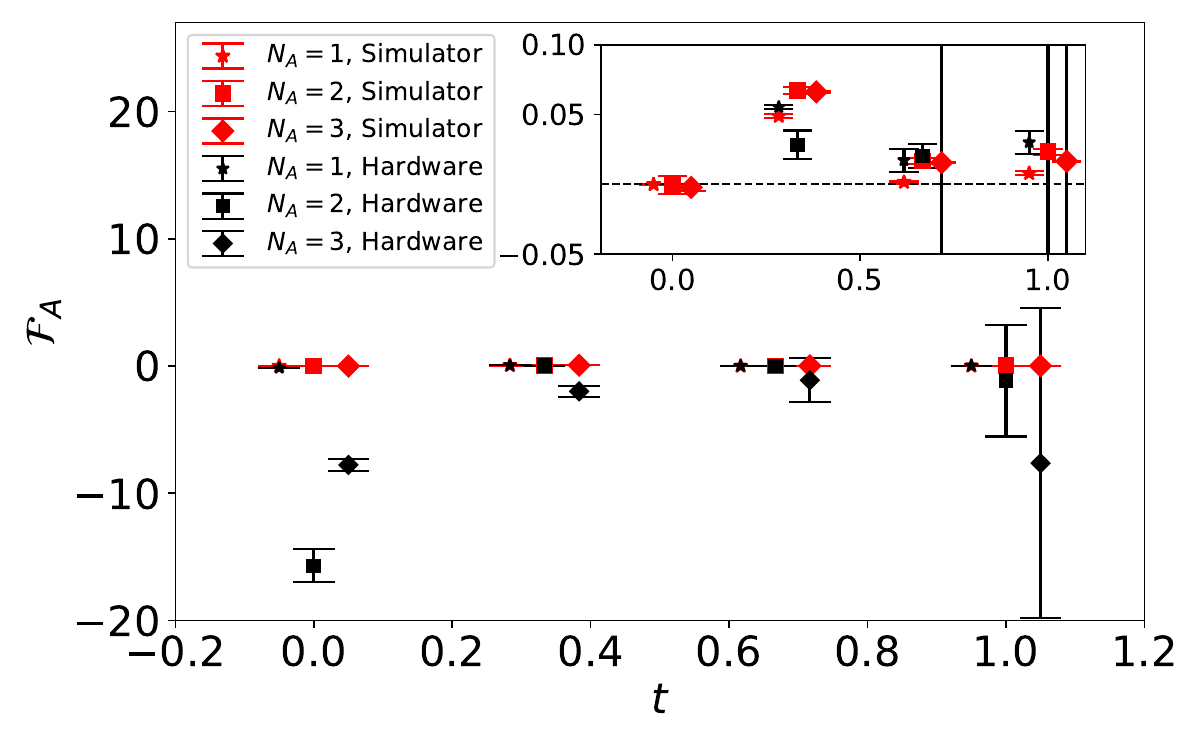}  %
}\hfill
\caption{R\'enyi-2 entropy and anti-flatness of entanglement spectrum as functions of time for system size $N=151$ and subsystem sizes $N_A=1$, $2$, and $3$. The notation follows that of the 
$N=101$ case shown in Fig.~\ref{fig:N101NA123}, except that the hardware experiments were carried out on the IBM quantum hardware ``{\it ibm\_kingston}" with 4,000 shots. 
}
\label{fig:N151NA123}
\end{figure*}

In contrast, the $N=151$ R\'enyi-2 entropy results shown in Fig.~\ref{fig:N151NA123} indicate that the hardware error accumulated in the quantum circuit exceeds the capability of the current error mitigation techniques when nearly all 156 qubits are used in the computation. The R\'enyi-2 entropies become strongly negative, even at $t=0$. This unphysical behavior may arise from large measurement errors on some qubits, as well as statistical and gate errors. These errors drive some eigenvalues of the reduced density matrix to negative values, while others are correspondingly inflated in order to preserve the unit trace condition.

\section{\JWX{Entanglement Spectrum and the Boltzmann Distribution}}
\label{app:Boltzmann}

\JWX{We examine whether the eigenvalues of the reduced density matrix, $\lambda_i$ follow a Boltzmann distribution, \XY{which is a prediction of the subsystem ETH~\cite{Dymarsky:2016ntg}}
\begin{equation}
\lambda_i \propto e^{-E_i/T},
\end{equation}
where $E_i$ denotes the corresponding subsystem energy. To reduce finite-volume effects, we estimate the subsystem energies by rescaling the energy eigenvalues of the full system according to the subsystem size. This is justified by the spatial homogeneity of our system, for which the energy density is uniform.}

\begin{figure}[!h]
\centering
  \includegraphics[width=0.44\textwidth]{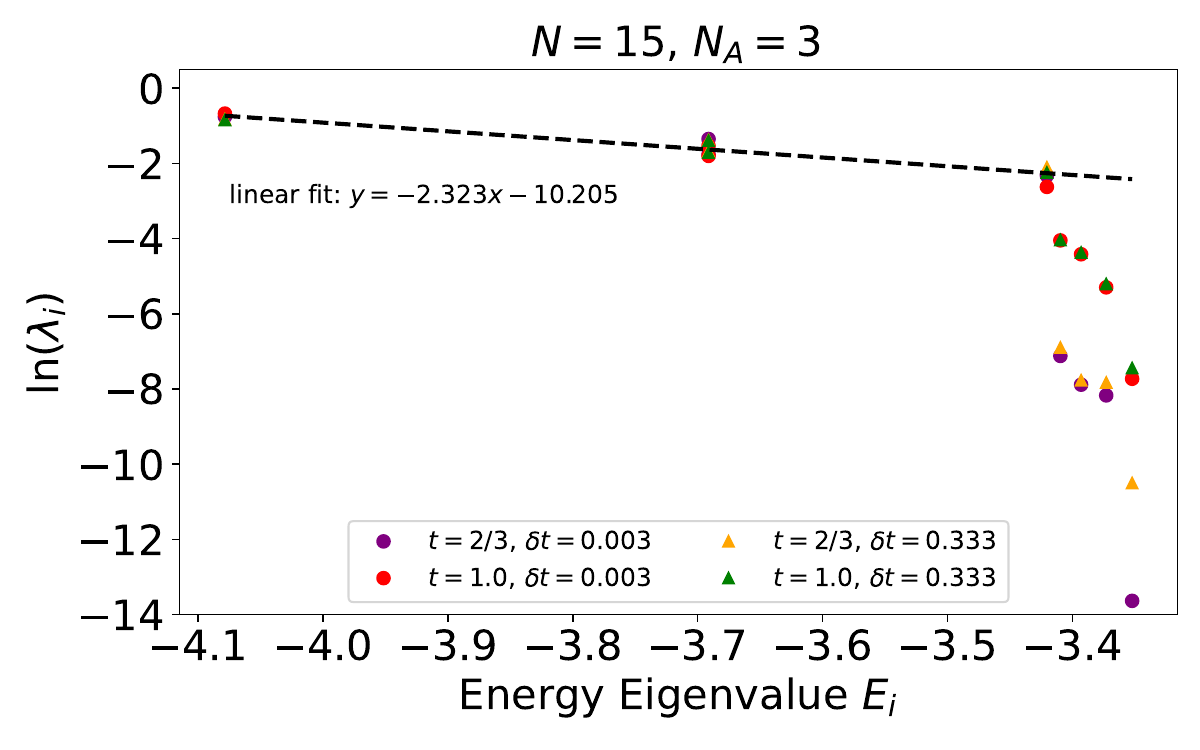} 
\hfill
\caption{\JWX{Logarithm of the spectral weights, $\ln(\lambda_i)$, as a function of the corresponding subsystem energies, $E_i$, for the lowest few energy states with $N=15$ and $N_A=3$. Results are shown for different Trotter step sizes $\delta t$.}
}
\label{fig:thermal}
\end{figure}

\JWX{The results are shown in Fig.~\ref{fig:thermal}, where $\ln(\lambda_i)$ is plotted against $E_i$ for the lowest few energy levels. In thermal equilibrium, the points are expected to lie on a straight line with slope $-1/T$ with temperature $T=0.431$. Because determining $E_i$ requires exact diagonalization, we carry out this analysis for the smaller system size $N=15$ using a classical simulator.}

\JWX{At $t=2/3$, the four lowest-energy states already lie approximately on a straight line, while the higher-energy states exhibit noticeable deviations. At $t=1$, the four lowest-energy states remain close to the same line, whereas the higher-energy states move progressively toward it. This indicates that, although the subsystem is not yet fully thermalized at $t=2/3$, the R\'enyi-2 entropy may already appear to have saturated. The remaining time dependence is mainly carried by eigenmodes with smaller spectral weights, whose contributions to the R\'enyi-2 entropy are correspondingly suppressed.}

\JWX{We further find that this behavior is largely insensitive to the Trotter step size $\delta t$. The main exception is the eigenmode with the smallest spectral weight, for which the relative error is naturally the largest.}

\newpage
\bibliographystyle{apsrev4-1}
\bibliography{Renyi_Entropy}

\end{document}